\documentclass[fleqn,usenatbib]{mnras}
\usepackage{newtxtext,newtxmath,mathtools}

\usepackage[T1]{fontenc}
\usepackage{ae,aecompl}
\usepackage{array,float}
\usepackage{comment}

\usepackage{graphicx}   
\usepackage{amsmath}    
\usepackage{scalerel}
\usepackage{xcolor}

\title{A search for the missing baryons with X--ray absorption lines towards the blazar \es}

\author[D. Spence et al.]{
David Spence$^{1}$, Massimiliano Bonamente$^{1}$\thanks{E-mail: bonamem@uah.edu}, Jukka Nevalainen$^{2}$, Toni Tuominen$^{2}$, 
\newauthor Jussi Ahoranta$^{3}$,   Jelle de Plaa$^{4}$,
 Wenhao Liu$^{5}$ and Nastasha Wijers$^{6}$\\
$^{1}$Department of Physics and Astronomy, University of Alabama in Huntsville, Huntsville, AL \\
$^{2}$Tartu Observatory,University of Tartu, T{\~{o}}ravere, Estonia\\
$^{3}$Department of University of Helsinki, Helsinki, Finland\\
$^{4}$SRON Netherlands Institute for Space Research\\
$^{5}$Purple Mountain Observatory, Chinese Academy of Sciences\\
$^{6}$Center for Interdisciplinary Exploration and Research in Astrophysics (CIERA) and Department of Physics and Astronomy,\\ Northwestern University, 1800 Sherman Ave, Evanston, IL 60201, USA\\
}
\date{Accepted XXX. Received YYY; in original form ZZZ}

\pubyear{2021}

\usepackage{color,nicefrac}
\def\E{\text{E}}

\def\hst{\it HST\rm}
\def\fuse{\it FUSE\rm}
\def\nodata{---}
\def\lya{Lyman-$\alpha$}
\def\nii{N~II}
\def\oi{O~I}
\def\oii{O~II}
\def\oiii{O~III}
\def\oiv{O~IV}
\def\ov{O~V}
\def\ovi{O~VI}
\def\ovii{O~VII}
\def\oviii{O~VIII}
\def\neviii{Ne~VIII}
\def\neix{Ne~IX}
\def\nex{Ne~X}

\def\civ{C~IV}
\def\hi{H~I}

\def\bla{BLA}
\def\nion{N_{\mathrm{ion}}}
\def\nionj{N_{\mathrm{ion,j}}}
\def\E{\mathrm{E}}
\def\chandra{\it Chandra\rm}
\def\xmm{\it XMM-Newton\rm}
\def\planck{\it Planck\rm}
\def\es{1ES~1553+113}
\def\kms{km~s$^{-1}$}
\def\nFUV{8}

\def\OWHIMX{\Omega_{\mathrm{WHIM,X}}}
\def\Nosevenplus{$N_{\mathrm{OVII}}^+$}
\def\Noseven{$N_{\mathrm{OVII}}$}
\def\eagle{\texttt{EAGLE}}
\def\spex{\texttt{SPEX}}

\def\cmin{$C_\text{min}$}

\begin{document}
\label{firstpage}
\pagerange{\pageref{firstpage}--\pageref{lastpage}}
\maketitle

\begin{abstract}
This paper presents an analysis of 
\xmm\ X--ray spectra of the quasar \es, in search for
absorption lines from the intervening warm--hot intergalactic medium.
A search for \ovii, \oviii\ and \neix\ resonance absorption lines was performed at eight fixed redshifts that feature \ovi\ or \hi\ broad Lyman--$\alpha$ absorption lines that were previously detected from HST data
The search yielded one possible
detection of \ovii\ at a redshift $z \simeq0.1877$ with an \ovi\ prior, with a statistical significance that is equivalent to a 2.6--$\sigma$ confidence level. 
The spectra were also stacked at the wavelengths of the expected redshifted \ovii\ and \oviii\ lines,  but the analysis did not reveal 
evidence for the presence of additional X--ray absorbing WHIM.
Moreover, the spectra were used to investigate two putative \ovii\ absorption lines that were detected serendipitously in an earlier analysis of the same data by F.~Nicastro and collaborators.
The paper also presents a comprehensive statistical framework for cosmological inferences from the analysis of 
absorption lines, which makes use of cosmological simulations for the joint probability distributions
of FUV and X--ray ions. Accordingly, 
%
%
we conclude that the new possible \ovii\ absorption at $z\simeq0.1877$
is consistent with a contribution from the hot WHIM to the baryon density in an amount of $\OWHIMX/\Omega_b = 44\pm22$\%. However, there are large systematic
uncertainties associated with the temperature and abundances of the absorbers, 
and only a larger sample of X-ray sources can provide an
accurate determination of the cosmological density of the WHIM.
\end{abstract}


\section{Introduction}
One of cosmology's open questions is the location of a fraction of the ordinary baryonic matter in the local universe. This issue, commonly referred to as the \textit{missing
baryons} problem, arises from the comparison between high--redshift detections
of baryons, primarily via the \lya\ forest, and measurements at low redshift,
whereas a sizable fraction of the high--redshift baryons appear to be
unaccounted for in the low--redshift universe \citep{shull2012}. 
For over two decades, numerical simulations have suggested that the
missing baryons may be located in filamentary structures of galaxies 
that host a warm--hot
intergalactic medium at temperatures of approximately $\log T(K)=5-7$ \citep[the WHIM, e.g.][]{cen1999,dave2001,bertone2008,cautun2014, martizzi2018,tuominen2021}.
The lower range of  WHIM temperatures has been effectively probed with
FUV absorption lines by \fuse\ and \hst, revealing a number of 
absorption lines in the spectra of background sources, primarily \hi\ broad Lyman--$\alpha$ (\bla),
\ovi, \civ, and other lines.
Such FUV absorption lines are likely to arise either from warm--hot gas in the the outer 
regions of individual galaxies or from the intergalactic WHIM, and they
track a significant fraction 
of the low--redshift baryons \citep[e.g.][]{tilton2012,shull2012,stocke2014,danforth2016}.
The higher range of WHIM temperatures have been more challenging to 
probe, primarily because the most prominent absorption lines associated with
a plasma at $\log T(K) \geq 6$ are in the X--ray range (i.e., the \ovii\ and \oviii\ resonance lines), and the available spectrometers on board
\xmm\ and \chandra\ are not as sensitive at the FUV instruments.

Despite the challenges associated with the flux of available background sources
and the resolutions and sensitivity of the X--ray spectrometers, 
a few detections of absorption lines from the WHIM have been reported,
including two in the spectrum of the quasar \es\ \citep{nicastro2018,nicastro2018b}, which is the subject of the present investigation. Other detections of the WHIM through
X--ray absorption include PKS~2155-304 \citep{fang2002,fang2007}, although the detection
was not confirmed by others \citep{yao2009, cagnoni2004, nevalainen2019}; Mrk~421 by \cite{nicastro2005}, which was however followed by reports of non--detections
by \cite{rasmussen2007} and \cite{yao2012}; in H~2356-309 by \cite{fang2010} and \cite{buote2009}; in Mrk~501 by \cite{ren2014}, 
in Ton~S180 \citep{ahoranta2021} and in 3C~273 \citep{ahoranta2020},
and finally in PG~1116+215 by \cite{bonamente2016,bonamente2019b}.
Finally, a tentative
detection in H~1821+643 was reported by \cite{kovacs2019} using stacking of signals at different redshifts, using a method
that is similar to the one used in this paper.

The study of EAGLE hydrodynamical simulations by \cite{wijers2019} suggests that there
is significant correlation between \ovi\ and \hi\ BLA FUV 
absorption lines and prominent X--ray ions such as \ovii\ and \oviii. This correlation
motivates our search for X--ray counterparts to FUV absorption lines in the spectrum 
of \es, and of other sources with FUV detections reported by the \hst\ and \fuse\ surveys
of \cite{tilton2012} and \cite{danforth2016}. In this paper we present the results of 
our search for \es, and develop the method of analysis that will be used for
a larger sample of FUV sources, including the use of systematic upper limits to the non--detection
of X--ray lines. 

This paper is structured as follows: in Section~\ref{sec:fuv} we describe our strategy for the search of X--ray absorption lines and the available FUV absorption data towards \es, in Section~\ref{sec:x}
the available X--ray data used for this investigation, in Section~\ref{sec:whim}
our analysis of the search and analysis of the  X--ray absorption lines. Finally,  Sect.~\ref{sec:cosmology} 
presents the cosmological implication of the results,  and
Section~\ref{sec:discussion} a discussion and the conclusions.

\section{The search for X--ray absorption towards \es}
\label{sec:fuv}

\cite{nicastro2018} conducted a blind search for X-ray
absorption lines in the \xmm\ spectrum of the X--ray bright blazar \es, and identified two absorption features interpreted as
\ovii\ He--$\alpha$ absorbers at $z \simeq 0.4339$ and 0.3557 possibly associated with the
WHIM towards the blazar. At those redshifts, there are a handful of galaxies
with confirmed redshifts that are at a projected distance of $\leq 1.5$~Mpc from the
sightline and with a velocity within 1,000~km~s$^{-1}$ of the X-ray redshifts.
These galaxies might be indicative of the presence of a filamentary structure that
may host WHIM gas, but no filament detection was attempted to confirm it.

\subsection{The redshift of \es}
\label{sec:redshift}
Unfortunately, the spectroscopic redshift of \es\ is unknown, because its  spectrum lacks 
intrinsic emission or absorption lines usually employed to measure it. 
Based on intervening \lya\  FUV absorption, \cite{danforth2010,danforth2016} constrained its redshift
to a range $0.413 < z < 0.56$, and possibly 
with a value of $z=0.49\pm0.04$, based on $\gamma$--ray observations \citep{abramowski2015}. These measurements were used by \cite{nicastro2018} to suggest a WHIM origin for the two serendipitous absorption lines they detected.

A recent study by \cite{dorigo2022} constrains the redshift of several of these
nearly featureless \emph{BL Lac}  blazars, which constitute a significant fraction of all blazars. The study
also includes new near-UV observations of \es\ that were not available at the time of the \cite{nicastro2018} paper. 
Based on the maximum redshift of the intervening \lya\ forest towards \es\ ($z=0.4131$), the  distribution of the difference between the spectroscopic redshifts and the maximum \lya\  redshift from other blazars (including the X--ray--bright 3C~273, PKS~2155-304 and Mkn~421) indicates a 95\% probability that the
redshift of \es\ is in the range $0.408 < z < 0.436$. Moreover, \cite{johnson2019} had shown that \es\ 
is likely a member of a group of galaxies located near $z = 0.433$. 

The combination of these results
suggests that the most likely redshift for \es\ is approximately $z \simeq 0.433$, in correspondence of
the strongest of the two \cite{nicastro2018} absorption lines. If this is correct, then the putative 
\ovii\ $z \simeq 0.4339$ absorption would be intrinsic to the source, and not associated with the WHIM.
For the main purpose of this paper, which is to investigate X--ray absorption lines 
at FUV--prior redshifts, the $z=0.4131$ \lya\  absorption towards \es\ sets a clear lower limit for the
redshift of the blazar, $z \geq 0.4131$, and therefore the search for absorption lines at redshifts that are \emph{lower} than this limit is warranted.
In Sects.~\ref{sec:whim} and \ref{sec:cosmology} we will further discuss the uncertainties associated with
the lack of a spectroscopic redshift for \es.

\subsection{The Search for the WHIM with FUV priors}
The search for WHIM X--ray absorption lines can be aided by
the presence of FUV absorption lines that may act as sign--posts for the 
higher--temperature ions, instead of performing a blind search at
all redshifts. The strategy we employ is to use \ovi\ or \hi\ \bla\ detections
to identify reliable absorption line systems, and then using the X--ray spectra
to search for possible counterparts (such as \ovii\ and  \oviii) associated with
the FUV ions, as pioneered by \cite{bonamente2016} and \cite{nevalainen2019}. The assumption of this search method is that the WHIM is multi--phase,
and that lower--temperature ions provide a prior for the redshift that 
increases the statistical significance and the overall credibility of any associated X--ray detection. 
This assumption is consistent with the positive detections we have previously reported
for 3C~273, Ton~S180 and PG~1116+215 \citep{ahoranta2020,ahoranta2021,bonamente2016}.
The use of FUV ions as indicators for 
\ovii\ and \oviii\ is also suggested by the analysis of EAGLE simulations
by \cite{wijers2019}, who finds significant correlation between both \ovi\ and
\hi\  absorption with both \ovii\ and \oviii. In particular,
the strongest correlation is between either the \neviii\ and \ovi\ FUV ions, and the \ovii\ X--ray ion,
while the correlation between \hi\ and X--ray ions is lower (see Fig.~14 of \citealt{wijers2019}). 

Our method consists of the analysis  of the following FUV absorption--line systems:\\
(a) \ovi\ detections where both lines in the $\lambda \lambda$ 
1031.9, 1037.6 \AA\ doublet are detected, in such a 
way that the identification of the FUV signal as the \ovi\ doublet is secure; and \\
(b) \hi\ \lya\ with broadening
$b \geq 40$~\kms, indicative of hydrogen at $\log T(K) \geq 5$ (as in \citealt{nevalainen2019}), which marks the typical boundary for WHIM 
gas.

Towards \es,
the relevant FUV absorption lines from the \cite{danforth2016} catalog are
reported in Table~\ref{tab:danforth}, for a total of \nFUV\ FUV absorption line systems
that meet our criteria as possible sign--posts for higher--energy X--ray ions. All of the FUV redshifts
are lower than the minimum source redshift discussed in the previous section ($z \geq 0.4131$).
The first two
detected \ovi\ systems are separated by $\Delta z=0.0002$, or a velocity of 60~km~s$^{-1}$, 
which  cannot be resolved by the X--ray spectrometers, and therefore they will be treated
as a single system for the sake of the X--ray search.
This is the same search method by which we identified two possible
X--ray WHIM systems in PG~1116+215 and in Ton~S180 \citep{bonamente2016,ahoranta2021}
that are counterparts of FUV detections,
in addition to lower--significance signals in the spectra of 3C~273 and PKS~2155-304 \citep{ahoranta2020,nevalainen2019}.

\section{X-ray data}
\label{sec:x}
This paper uses data from the \xmm\ RGS and \chandra\ LETG grating spectrometers.
The methods of analysis of the X--ray data follows those presented
in \cite{nevalainen2017} and \cite{ahoranta2020} for \xmm, and in \cite{bonamente2016} for
\chandra. In the following we summarize the main features of the
data and data processing.

\subsection{Processing}
\subsubsection{\xmm/RGS}
We analyzed all RGS observations of 1ES1553+113 available as of May 2022,
for a total of 23 observations~\footnote{A  short observation (ObsID: 081083701) with exposure of $\sim8$ ks is discarded due  to incomplete observation files.} with total RGS exposure time of $\sim1.8$ Ms. The observations
are reported in Table~\ref{tab:Data}.

\begin{table}
    \caption{Log of \xmm\ and \chandra\ observations of \es}
    \label{tab:Data}
    \centering
    \begin{tabular}{ll}
        \hline
        \hline
        Obs ID &  Exp. time (s) \\
        \hline 
        \xmm\ & \\
0094380801 & 4374\\
0656990101 & 21815\\
0727780101 & 33307\\
0727780201 & 35001\\
0727780301 & 28605\\
0727780401 & 28701\\
0727780501 & 28190\\
0727780601 & 25482\\
0761100101 & 136312\\
0761100201 & 132836\\
0761100301 & 137321\\
0761100401 & 131393\\
0761100701 & 88618\\
0761101001 & 131870\\
0790380501 & 96482\\
0790380601 & 102300\\
0790380801 & 113713\\
0790380901 & 129190\\
0790381001 & 91261\\
0790381401 & 123553\\
0790381501 & 138104\\
0810830101 & 31039\\
0810830201 & 28360\\
         \hline 
         Total & {1,817,827}\\
         \hline
         \hline
         \chandra & \\
         12915 & 166307\\
         12916 & 153935\\
         12917 &175403 \\
         \hline
         Total & 495,645 \\
         \hline
         \hline
    \end{tabular}
\end{table}

 We used the \xmm\ SAS 18.0.0 software with the most recent calibration files available on July 2020 (XMM-CCF-REL-378) for processing the first order data. The data were reduced with the \texttt{rgsproc} pipeline using mostly the standard parameter values, except what is discussed below. In general, we follow the processing procedure laid out in \cite{ahoranta2020}.
Radiation damage caused by cosmic rays induces local enhancement of dark current, or extra charge traps \citep{devries2014}, which manifest themselves as hot and cool pixels (i.e. bad pixels). Permanent bad pixels have been mapped by the RGS team and the information on their location etc. is included in the calibration files. The standard processing with default parameters uses this information to reject the data from hot pixels but not from the cool pixels. Following \cite{ahoranta2020}, we employed the recent option \texttt{keepcool=no} to also reject the data from cool pixels. 
In addition, time-variable bad pixels not rejected by the standard pipeline may affect the given observation, and they were removed upon visual inspection of the spectra. 


The spectra were extracted with a 10~m\AA\ bin size, and then rebinned to 20~m\AA\ for the analysis.
We then co-added the first-order spectra from different observations, keeping RGS1 and RGS2 separate, using the \texttt{rgscombine} procedure. These co-added spectra were then converted 
into the \texttt{SPEX\footnote{https://www.sron.nl/astrophysics-spex}} format using the \texttt{trafo} (version 1.04) software.

\subsubsection{Chandra/LETG/HRC}

The \chandra\ data were processed with CIAO 4.11 using the
standard processing pipeline 
\texttt{chandra\_repro}. The 3 HRC observations available
for \es\ are listed in Table~\ref{tab:Data}. For each 
observation, we combined the $\pm1$ order LETG grating spectra
into a single spectrum. The spectra were then combined
into a single spectrum for analysis with SPEX using the
same \texttt{trafo} software used for the analysis of 
\xmm\ data. Additional details on the data reduction methods
are in \cite{bonamente2016}.

\subsection{Sensitivity of the data for the detection of absorption lines}
\label{sens.sec}
Before embarking on a study of a large number of relatively weak absorption lines,
we investigated  the constraints set by statistical and systematic uncertainties.
The RGS1 and RGS2 1st order effective area calibration uncertainties have been estimated as $\sim$2\% in the 10--30 \AA\ band, except for certain regions (such as near the oxygen edge) where 3-5\% uncertainties 
may occur \citep{kaastra2018}. The uncertainties in the calibration of the effective
area for the LETG/HRC instruments have been estimated at $\leq 10$\% across the entire band.\footnote{See https://cxc.cfa.harvard.edu/cal/Letg/LetgHrcEA/}

In order to evaluate the effect of possible effective area calibration inaccuracy on our results, we approximated the effect as a 2\% dip in the predicted continuum flux within a bin of 60 mÅ size, i.e. the RGS resolution element, relative to the surrounding continuum. This is intended 
as a means to model a typical scenario in which a mis--calibrated effective area leads to
a spurious absorption line--like feature.
We then modelled this feature with a Gaussian line model (\texttt{line} in SPEX) 
with a broadening of  100 km/s at a nominal wavelength of 20 \AA, and 
found that this feature requires a value of the optical depth 
parameter $\tau_0$= 0.37, producing an absorption line with equivalent width of   
EW$_\mathrm{sys}$ = 2.7 mÅ. This corresponds to $\log N_{\text{ion}}$ (cm$^{-2}) \simeq 15.0$,
using a \texttt{slab} model for an He--$\alpha$ \ovii\ absorption line. This is an indication
that typical systematic uncertainties do not permit the detection of
\ovii\ column densities less that approximately $\log N_\mathrm{OVII}$ (cm$^{-2}) = 15.0$.  For the \chandra\ HRC data, a 10\% uncertainty in the effective area in a resolution element of $\sim$60~m\AA\ corresponds to
a value of  $\tau_0$= 1.1, and it results in the inability to detect
absorption lines with below EW$_\mathrm{sys}$ = 14 m\AA. In turn, this corresponds
to an \ovii\ column density of  $\log N_{\text{ion}}$ (cm$^{-2}) \simeq 15.7$, 
following the same analysis as for the \xmm\ data.

\begin{figure}
\includegraphics[width=3.5in]{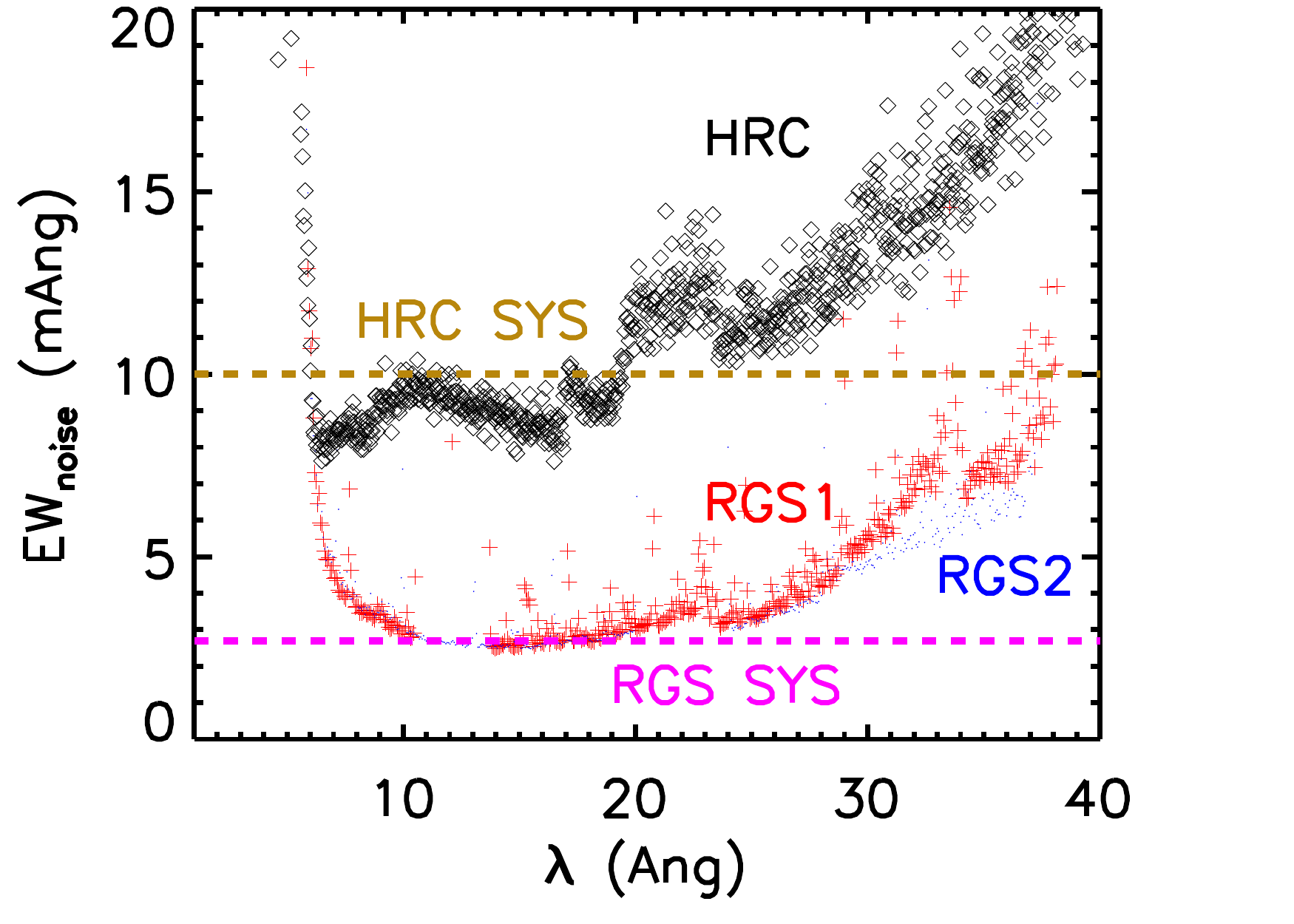}
\caption{The noise level in terms of the equivalent width of an absorption line for RGS1 (red), RGS2 (blue) and HRC (black). The limits set by the effective area systematics for RGS and HRC are indicated with a purple and brown dashed lines, respectively.
}
\label{sens.fig}
\end{figure}

We then calculated the relative statistical uncertainties of the RGS and
HRC spectra, using the background-subtracted count rate and its statistical uncertainty within each resolution element across the full waveband of the data. We then converted the relative statistical uncertainties into the equivalent width
of a spurious line allowed by the statistical uncertainties scaling linearly (2\% relative error corresponding to 2.7 mÅ as above), see Fig \ref{sens.fig}.
This calculation provides two useful results.
First, the number of photons is so high that the level of the statistical uncertainties of the data is comparable to that of the systematic uncertainties. 
Second, the \chandra\ HRC/LETG data are significantly less
sensitive than the RGS data, primarily due to their lower exposure time. We therefore
do not use the \chandra\ data in the detection
of possible absorption lines from the WHIM, and for this task
we focus on the more sensitive \xmm\ RGS spectra. The \chandra\ data
are analyzed in Sect.~\ref{sec:chandra} in order to assess the consistency of the
\xmm\ results with the lower quality \chandra\ data. The background levels of both the \chandra\ and the \xmm\ are also reported in Table~\ref{tab:background}, in two representative wavelength ranges.
Possible sources of systematic errors are described in Sect.~\ref{sec:systematics}.

\begin{table}
\centering
    \caption{Background levels near two characteristic wavelengths corresponding to the
    He~$\alpha$ line from \ovii, for the \xmm\ and \chandra\ spectra.}
    \label{tab:background}
     \begin{tabular}{c | c c | c c}
 \hline\hline
\ovii\ He$\alpha$  & \multicolumn{2}{c}{XMM} & \multicolumn{2}{c}{Chandra LETG}\\
        z & RGS1  &  RGS2  &  +1 order&  -1 order\\
 \hline
$0.1876$ &8.2\%  &7.7\% & 27.2\% & 30.8\%\\
$0.4339$ &34.5\%  &13.9\%  & 35.2\% & 43.4\%\\
\hline 
\end{tabular}
\end{table}


\section{Analysis of X--ray lines from the WHIM}
\label{sec:whim}


\subsection{WHIM ions and Galactic lines}
\begin{figure}
    \centering
    \includegraphics[width=3.0in]{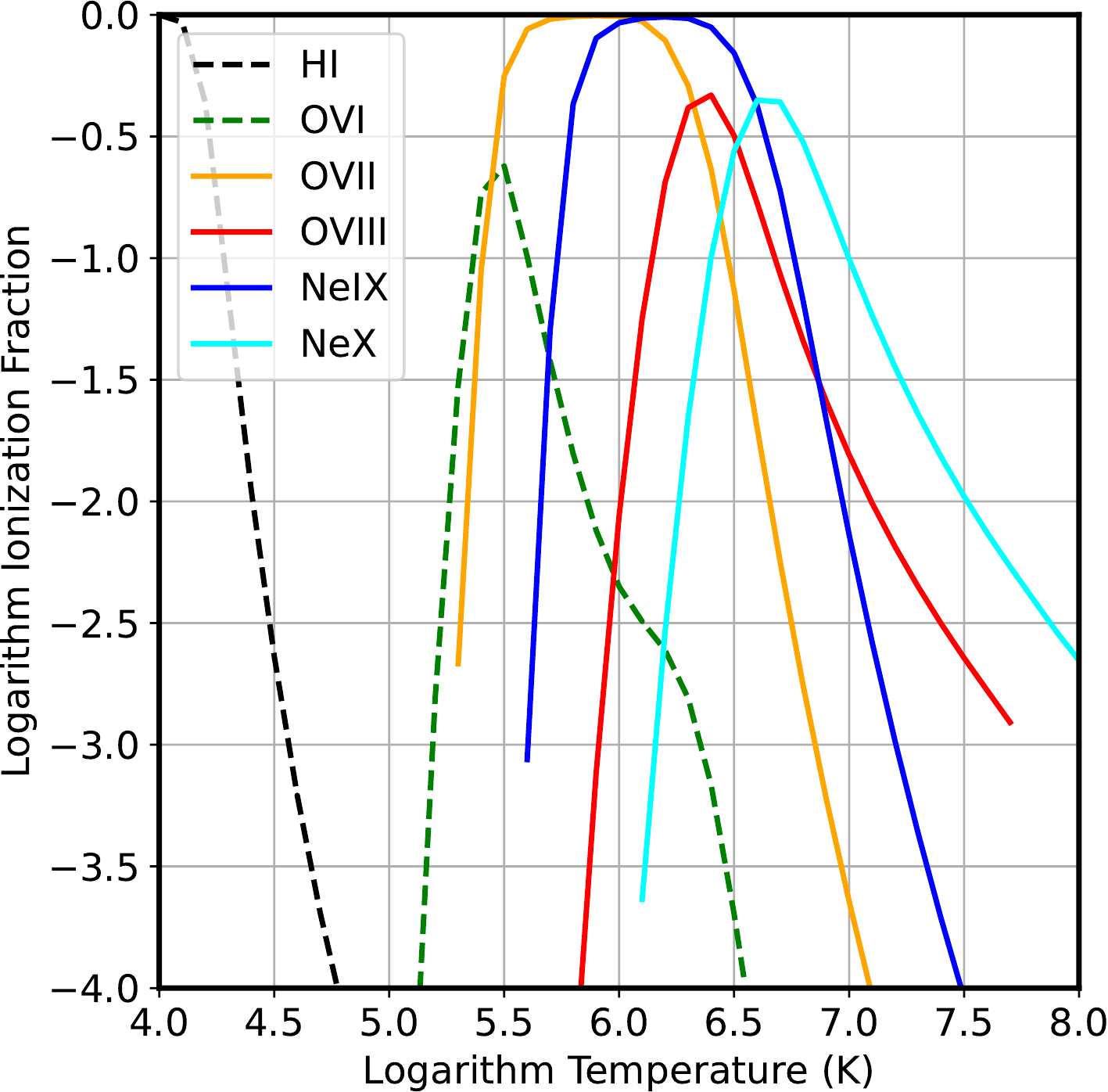}
    \caption{Ionization curves of key ions from \protect\cite{mazzotta1998}.}
    \label{fig:ions}
\end{figure}

In collisional ionization equilibrium (CIE) and for metal abundances 
proportional to the Solar ratios \citep[e.g.][]{anders1989}, 
the absorption lines with the largest expected equivalent width 
at a peak temperature $\log T(K) \geq 6$ are the
Lyman--$\alpha$ \oviii, He--$\alpha$ \ovii, 
and the \neix\ He--$\alpha$ and Lyman--$\alpha$ \nex\ (see Table~\ref{tab:Lines}).
The ionization curves of key ions of interest to this study are shown in Figure~\ref{fig:ions} \citep{mazzotta1998}.

Of particular interest to the detection of the WHIM in the $\log T(K) =6-7$ temperature
range are the \ovii\ and \oviii\ ions, which are expected to provide the most
prominent absorption line features in the \xmm\ grating spectra.
One of the main sources of possible confusion and misidentification for redshifted WHIM lines
are absorption lines from lower--ionization X--ray oxygen ions that generate from inner--shell transitions. In recent years there has been significant theoretical
and experimental progress in the identification of inner--shell transitions in oxygen ions
\citep[e.g.][]{gatuzz2015,gu2005,garcia2005}, with some of the strongest 
relevant absorption lines
from inner--shell oxygen ions being reported in Table~\ref{tab:Lines}.
The net effect of these oxygen--series lines is to make the wavelength range
from 21.6~\AA\ (the $z=0$ \ovii\ He-$\alpha$ line)  to 23.51 \AA\ (the 
$z=0$ \oi\ $1s-2p$ line, marking the location of the oxygen edge) effectively unavailable to search for \ovii\ and \oviii\ absorption
lines from the WHIM, given possible velocity structure in the Galaxy, and the resolution
of the grating spectrometers.

\begin{table*}
    \centering
    \caption{Wavelengths and atomic properties of key WHIM lines,
    and of inner--shell oxygen Galactic lines that may cause confusion in the identification of redshifted oxygen 
    WHIM lines.
    References for atomic data: \protect\cite{verner1996} (V96); \protect\cite{gu2005} (G05); \protect\cite{gatuzz2015} (G15);  \protect\cite{gharaibeh2011} and \protect\cite{nicastro2018b} (G11). See also Table~1 of \protect\cite{nevalainen2017} for other lines of relevance to the WHIM, and \protect\cite{garcia2005} for other $K\alpha$ atomic data for oxygen.} 
    \label{tab:Lines}
    \begin{tabular}{lllll}
    \hline
    Ion & Line & Wavelength (\AA) & Osc. strength  & Reference\\
    \hline
    \multicolumn{5}{c}{WHIM lines}\\
    \hline
         \ovii & $1s^2-1s2p$ (He$\alpha$) & 21.602 & 0.696  &  V96\\
         \ovii & $1s^2-1s3p$(He$\beta$) & 18.629 & 0.146 & V96 \\
         \oviii & $1s-2p$ (Ly$\alpha$)& 18.969 (18.973,18.967) & 0.416 & V96\\
         \oviii & $1s-3p$ (Ly$\beta$) & 16.006 (16.007,16.006) & 0.079 & V96 \\
         \neix & $1s^2-1s2p$ (He$\alpha$) & 13.447 & 0.724 & V96\\
         \nex & $1s-2p$ (Ly$\alpha$) & 12.134 (12.138,12.132) & 0.416  & V96\\
         \hline
         \hline
         \multicolumn{5}{c}{Galactic lines}\\
         \hline
         \oi & $2p^4-[1s]2p^5$  & 23.506 & \nodata &  G15\\
         \oi & $2p^4 - [1s]2p^4 3p$  & 22.889 & \nodata & G15\\
         \oii & $2p^3-[1s]2p^4$ & 23.346 & \nodata &  G15\\
         \oii & $2p^3 - [1s]2p^33p$ & 22.287 & \nodata & G15 \\
         \oiii & $2p^2-[1s]2p^3$ & 23.110, 23.057 & \nodata & G15  \\
         \oiv & $1s^22s^22p-1s2s^22p^2$ & 22.741 & 0.46 &  G05\\
         \ov & $1s^2 2s^2-1s2s^22p$ & 22.363 & 0.55 & G05\\
         \ovi & $2s^2 - [1s]2s2p$ & 22.024 & \nodata & G15 \\
         \nii &  $1s^2 2s^2 2p^2 -1s 2s^2 2p^3$ & 29.99-31.16 & \nodata & G11\\
         \hline
    \end{tabular}
\end{table*}

\begin{table*}
\caption{List of \ovi\ and \hi\ \bla\ lines in the FUV spectra of \es\ 
from \protect\cite{danforth2016}. The numbers identify absorption line systems 1--8}
\label{tab:danforth}
\begin{tabular}{llccccc}
\hline
&Redshift &  Obs. wavelength $\lambda$ (\AA)  & Line ID & EW (m\AA) &  $b$ (km~s$^{-1}$) & $\log N$ (cm$^{-2}$)\\
\hline
1 & 0.187601 & 1225.52 & OVI 1032 & 64$\pm$5 & 14.9$\pm$2.5 & 13.85$\pm$0.09 \\
2 &0.187774 & 1225.69 & OVI 1032 & 65$\pm$8 & 15.3$\pm$3.5 & 13.85$\pm$0.11\\
3 & 0.189833 & 1227.82 & OVI 1032 & 37$\pm$37& 25.0$\pm$4.9 & 13.51$\pm$2.18\\
1 & 0.187570 & 1232.24 & OVI 1038 & 24$\pm$5 & 6.5$\pm$4.0 & 13.71$\pm$0.08\\
2 & 0.187705 & 1232.38 & OVI 1038 & 50$\pm$6 & 39.6$\pm$6.5 & 13.94$\pm$0.06\\
3 & 0.189858 & 1234.62 & OVI 1038 & 10$\pm$3 & 20.1$\pm$9.9 & 13.28$\pm$0.22\\
4 & 0.394964 & 1439.52 & OVI 1032 & 69$\pm$7 & 37.5$\pm$2.6 & 13.80$\pm$0.04\\
4 & 0.394995 & 1447.46 & OVI 1038 & 64$\pm$12& 57.24$\pm$.8 & 14.04$\pm$0.07\\
\hline
5 &0.034656 & 1257.78 & Lya 1215 & 69$\pm$7  & 72.0$\pm$9.4 & 13.12$\pm$0.04\\
6 &	0.042726 & 1267.59 & Lya 1215 & 117$\pm$6 & 63.0$\pm$4.5 & 13.38$\pm$0.02\\
7 &	0.063637 & 1293.01 & Lya 1215 & 47$\pm$23 & 76.3$\pm$19.9& 12.96$\pm$0.16\\
8 &	0.218690 & 1481.50 & Lya 1215 & 28$\pm$8 & 62.6$\pm$21.8 & 12.72$\pm$0.12\\
\hline
\end{tabular}
\end{table*}

\begin{table*}
\caption{Results of the spectral fits to the 1ES1553+113 \xmm\ data with the \texttt{spline} and \texttt{slab} models.}
 \label{tab:results}
 \renewcommand{\arraystretch}{1.2}
 \hspace*{-0.75cm}
 \begin{tabular}{l c c c c c c c c c c | l}
 \hline\hline
 \multicolumn{3}{c}{$z$} & \multicolumn{2}{c}{$C$ stat.} & \multicolumn{2}{c}{\ion{O}{VII} }  & \multicolumn{2}{c}{\ion{O}{VIII} } & \multicolumn{2}{c}{\ion{Ne}{IX}} & $\Delta C$ (dof)\\
        \# & Value  & Fixed/Free & Value  & d.o.f. & $\log{N}$(cm$^{-2}$) & $\Delta\ C$  & $\log{N}$(cm$^{-2}$)  & $\Delta\ C$ & $\log{N}$(cm$^{-2}$) & $\Delta\ C$ & \\
 \hline
\multicolumn{10}{c}{Redshift priors from FUV}\\
1 & $0.1876$ & Fixed & 1861.8 & 1478 & 15.13$^{+0.19}_{-0.31}$ & 4.2 & 15.3$^{+0.3}_{-8.3}$ & 0.8  & 15.26$^{+0.26}_{-0.62}$ & 1.8 & 6.8 (3)\\
2 & $0.1878$ & Fixed & \multicolumn{8}{c}{\nodata} \\
3 & $0.1898$ & Fixed & 1865.8 & 1478 & 9.3$^{+5.3}_{-2.3}$ & 0.0 & 15.50$^{+0.24}_{-0.49}$ & 2.3 & 15.26$^{+0.26}_{-0.66}$ & 1.7 & 4.0 (3)\\
4 & $0.3950$ & Fixed & 1869.6 & 1478 & 14.7$^{+0.4}_{-7.7}$ & 0.3 & 14.0$^{+1.1}_{-7.0}$ & 0.0 & 8.0$^{+6.3}_{-1.0}$ &  0.0 & 0 (3)\\
5 & $0.0347$ & Fixed & 1869.7 & 1478 & 14.5$^{+0.6}_{-7.5}$ & 0.1 & 13.6$^{+1.5}_{-6.6}$ & 0.1  & 7.0$^{+8.0}_{-0.0}$ & 0.1 & 0.1 (3)\\
6 & $0.0427$ & Fixed & 1869.0 & 1478 & 14.9$^{+0.4}_{-7.9}$ & 0.5 & 15.0$^{+0.4}_{-8.0}$ & 0.8  & 7.0$^{+7.4}_{-0.0}$ & 0.7 & 0.6 (3) \\
7 & $0.0636$ & Fixed & 1869.1 & 1479 &  \multicolumn{2}{c}{\nodata}& 9.4$^{+5.7}_{-2.4}$ & 0.0 & 15.1$^{+0.4}_{-8.1}$ & 0.7  & 0.7 (2) \\
8 & $0.2186$ & Fixed & 1867.9 & 1480 & 14.97$^{+0.24}_{-0.57}$ & 2.0 &  \multicolumn{2}{c}{\nodata} & \multicolumn{2}{c}{\nodata} & \nodata\\
\hline 
\multicolumn{10}{c}{Redshifts from \protect\cite{nicastro2018}}\\
9 & $0.4336^{+0.0005}_{-0.0003}$ & Free & 1835.6 & 1477 & 15.74$\pm{0.11}$ & 30.9 & 14.2$^{+0.9}_{-7.2}$ & 0.0 & 15.30$^{+0.25}_{-0.38}$& 2.6 & 34.1 (4) \\ 
10 & $0.3557^{+0.0006}_{-0.0006}$ & Free & 1864.3 & 1477 & 15.29$\pm{0.22}$ & 4.8& 9.9$^{+5.7}_{-2.9}$ & 0.0& 14.0 $^{+0.3}_{-7.0}$ & 0.7 & 5.5 (4) \\ 

\hline
\hline
\end{tabular}
\end{table*}

\subsection{Models for continuum emission and absorption lines}

The analysis of the \xmm\ spectra was limited to the 13--33~\AA\ range, where all lines of interest
from the \ovii\, \oviii\ and \neix\ ions are located, at the redshifts provided by the FUV priors.
The continuum emission in the \es\ spectra were fit to a cubic spline model
(\texttt{spline} in SPEX), given that a simple power--law model across the 
entire range of interest does not provide a satisfactory fit, due to the 
relative large count rate of the source. 
Foreground absorption by the Galaxy was modeled with two \texttt{hot} model components,
one for the neutral disk and one for the hot halo, same as in the \cite{ahoranta2020} investigation of 3C~273, to which the reader is referred for further details.

Possible absorption by the WHIM is modeled with a \texttt{slab} model, which provides
the atomic data and the column density of all possible ions of interest, including
\ovii, \oviii, and \neix. This model has the advantage of allowing for \emph{all} known absorption lines from the chosen ions to be modeled simultaneously. The model does not enforce a specific temperature or method of ionization for the absorbing plasma (as is
the case for the \texttt{hot} model), so that all 
possible absorption line features can be identified and studied.
This is the same WHIM absorption model used in previous studies, such as
\cite{nevalainen2017} and \cite{ahoranta2020}.


\subsection{Analysis and results for lines with FUV priors}
\label{sec:FUVpriors}
\subsubsection{Methods of spectral fitting}
A \texttt{slab} WHIM absorption model was used to investigate the presence of absorption lines at
the eight redshift systems identified from the FUV data (see Table~\ref{tab:danforth}).
For each of the 24 redshifted ions identified in the FUV (the 8 systems from
Table~\ref{tab:danforth}, each with three possible X--ray ions), the \texttt{spline} 
model was supplemented by one \texttt{slab} model at a time,  and the  
constraints on X-ray absorption for these FUV systems are provided in the top part of
 Table~\ref{tab:results}.
  Redshift systems 1 and 2 are separated in redshift by such a small amount that a
 redshifted \ovii\ K$\alpha$ lines would be separated by just 4~m\AA, well below the resolution of these data.
 Therefore we did not repeat the fits for redshift system 2, and the two redshift systems share the same X--ray analysis.
 
 Specifically, we performed 7 different fits, where the same \texttt{spline} continuum 
 model (and the foreground model components) was supplemented by a different \texttt{slab} model in each fit,
 in which the \ovii\, \oviii\ and \neix\ column densities were left free to vary, at a fixed
 redshift. In so doing, the fit is able to determine whether there is significant absorption from any
 of the ions under consideration, at that fixed redshift, and from all the ions as a whole. 
 The results of these fits are reported in Table~\ref{tab:results},
 where each line corresponds to each of the seven regressions at a fixed value of the redshifted absorbing material.
 
 The possible \ovii\ at redshift system 6 ($z=0.0427$) falls
 at a redshifted wavelength of $\lambda=22.524 $, which is nearly indistinguishable from
 the redshifted wavelength of \oviii\ at redshift system 1-3 ($\lambda=22.528-22.569$), and therefore
 some of these fits overlap the same wavelengths. For some ions, such as \ovii\ at redshift system 7 and \neix\ at redshift
 system 8, the main expected absorption lines falls in or near a gap in the data. Accordingly, since 
 these data could not constrain them effectively, the result of the fit
 was not reported.
 
 \subsubsection{Statistical methods for the WHIM absorber component}
 The main goal of these spectral models is to determine the presence of X--ray absorbing material
 along the sightline and at a fixed redshift. This, in turn,
 means determining whether the additional \texttt{slab} component is 
 significant in the fit. 
 To aid with the determination of the  significance of detection of a specific ion at
 a fixed redshift, the $\Delta C = C - C_{\text{min}}$ statistic is also reported
 in the same table, to represent the increase in the $C$ statistic when the
 additional \texttt{slab} component is ignored. To measure the $\Delta C$ statistic, we fixed
 the column density of each of the three ions at the lowest value allowed by the model 
 (which is $\log N (\text{cm}^{-2})=7$, de facto corresponding to zero column density), 
 and repeated the fit
 in order to calculate the increase in the fit statistic with one fewer model parameter. We
 also performed an additional fit in which the column density of all three ions were simultaneously
 frozen to the lowest value, and calculated the overall $\Delta C$ statistic with 3 fewer free parameters. This is
 the value reported in the rightmost column of the table, along with the number of additional degrees of freedom in the fit. 
 For redshift system 7 and 8, some of the lines
 were unobservable, thus the number of free parameters in the fit was adjusted accordingly.
 
 The $\Delta C$ statistic is a likelihood--ratio statistic,
as described  by S.S.~Wilks \citep{wilks1938,wilks1943} and widely
 used for astronomical statistics \citep[e.g.,][]{cash1979,protassov2002}.
 Under a number of mathematical conditions described in detail in \cite{cramer1946},
 a likelihood--ratio statistic such as $\Delta C$ (but also, for example, the $\Delta \chi^2$ 
 or $F$ statistics),
 is asymptotically distributed like
 the $\chi^2$ distribution, for sources with sufficiently large number of counts 
 \citep{kaastra2017,bonamente2020}, as is
 the case for these data. The number of degrees of freedom of the parent $\chi^2$ distribution is
 determined by the number of free parameters of the additional model component.
  Since the column density of the ion is the only free parameter
 in the fit for the $\Delta C$ statistics reported in Table~\ref{tab:results} (except those in
 the last column, where
 the number of degrees of freedom is reported explicitly), the null hypothesis that there is no absorption
 from that ion yields a parent distribution for the statistic that
 is approximately $\Delta C \sim  \chi^2(1)$, with 90\%, 99\% and 3--$\sigma$ (99.7\%) critical values
 of respectively $C_{\text{crit}}=2.7, 6.6, 8.8$.
 
 Before interpreting the results of Table~\ref{tab:results}, it is necessary to remember that
 one of the conditions for the use of this likelihood--ratio statistic is that the
 additional component is \emph{nested}, i.e., it can be zeroed--out by a suitable
 choice of the additional parameter(s), corresponding to the
 null hypothesis that there is no absorbing material. The other main condition is that the
 value of the parameter(s) that correspond to the null hypothesis is \emph{not} on the boundary
 of the allowed parameter space. While the first condition is clearly satisfied, the
 additional \texttt{slab} model component fails the second condition, in that the absorber
 can only have a positive column density, i.e., it can only give rise to an absorption line,
 and not an emission line. This issue is described in detail in \cite{protassov2002}, including
 an indication that often (but not always or necessarily) using the $\chi^2(1)$
 distribution as a test for the measured $\Delta C$ leads to a \emph{conservative}
 test of the necessity of the component, when the null hypothesis value of the parameter
 falls at the boundary of the allowed parameter space. This means that, if the measured $\Delta C$
 exceeds the critical value of the $\chi^2(1)$ distribution, the component should be regarded 
 as significant at that level of confidence, or higher.~\footnote{The situation is illustrated by \cite{protassov2002} considering that, for a value of the parameter on the boundary, say
 $x=0$ and with only positive values allowed, all possible positive values of the parameters
 would default to 0 in the fit with a dataset that follows the null hypothesis.
 In turn, the distribution of the likelihood--ratio fit statistic would feature a delta function
 at $x=0$ with approximately \nicefrac{1}{2} probability, and the remainder of the positive values
 would feature a distribution that is approximately $\chi^2$, only with reduced normalization. Thus,
 critical values of the $\chi^2$ distribution are larger than those of this hybrid distribution,
 thus leading to a conservative test for the rejection of the null hypothesis.}
 
 \subsubsection{Analysis with a simplified power--law model}
 
 Given that the requirements for testing the null hypothesis with a likelihood--ratio test
 are not fully satisfied by the \texttt{slab} model, we also performed additional fits
 consisting of the use of a simple power--law emission model in a narrow wavelength range,
 supplemented by a
 phenomenological \texttt{line} model, which allows for \emph{both} negative and positive
 fluxes around the null hypothesis of an optical depth at line center of $\tau_0=0$, thus
 satisfying the criteria for hypothesis testing with a $\chi^2$ distribution for the
 resulting $\Delta C$ statistics.
 Although this model does not have the convenient features of the \texttt{slab} model (i.e.,
 the atomic physics needed to interpret a fluctuation as an ion column density), 
 this additional regression is more readily capable of answering the question of whether
 there is a line--like fluctuation in correspondence of the expected X--ray lines.
 These additional fits were only performed for a few selected lines where the results of
 Table~\ref{tab:results} indicates the possibility of an absorption line feature. They
 are summarized in Table~\ref{tab:po}, where the $p$ values correspond to the use of the $\chi^2$ distribution for the $\Delta C$ statistics. The \texttt{line} model was used as a Gaussian profile 
 (setting to zero the Lorentzian component), and with
 a fiducial line width fixed at 10~m\AA, since its use is only to detect the presence of
 a fluctuation at the target wavelength.
  Relevant portions of the spectra are shown in Figure~\ref{fig:z0.1876} for the \ovii, \oviii\
 and \neix\ ions at $z=0.1876$ (redshift 1).
 In Table~\ref{tab:po}, column densities associated with the line component are estimated
 via the equivalent width of the line provided by \texttt{SPEX} as part of the fit, and with the assumption
 of an optically thin line \citep[e.g., using Eq.~2 of][]{bonamente2016}.

 \begin{table*}
 \caption{Additional fits using a power--law continuum model and a \texttt{line} model. For each
 regression, a band of 1~\AA\ around the expected line center was used in the fit. For the lines
 with FUV prior, the line center was fixed. For the other lines from \protect\cite{nicastro2018},
 the line center was left free to vary, and the index of the power-law was fixed at a fiducial value.
 The last two column report the redshift--trial--corrected $P$--values, according to the binomial 
 distribution method, and using a simulation following the \protect\cite{kaastra2006} method. For the \texttt{line} model,
 the implied column densities are estimated from the measured equivalent widths, assuming optically thin lines
 (i.e., Eq.~2 of \protect\citealt{bonamente2016}).
  }
 \label{tab:po}
 \begin{tabular}{c|llllll|llll}
 \hline
 \hline
 Target line & \cmin& \multicolumn{2}{c}{power--law component} & \multicolumn{3}{c}{line component} &
                $\Delta C$& $p$--value & \multicolumn{2}{c}{Corrected $p$--value}\\
             & (d.o.f.)       & \text{norm.} & \text{index} & $\lambda$ (\AA) & $\tau_0$ & $\log N (\text{cm}^{-2})$ & (d.o.f.)& & $P_{\text{binom}}$ & $P_{\text{sim}}$\\
 \hline 
\ovii\ $z=0.1876$ (\# 1) & 103.5(87) & 1968$\pm^{685}_{481}$ & 0.886$\pm^{0.385}_{0.412}$ & 25.6545 
& $0.57\pm^{0.27}_{0.25}$ & $15.26\pm^{.14}_{.21}$ & 6.6(1) & $1.0\times 10^{-2}$ & \nodata{} & \nodata{}\\[5pt]
                         &        & 1909$\pm^{661}_{463}$ & & & & & & \\[5pt]
\oviii\ $z=0.1898$ (\#3) & 38.4(28) & $760.9\pm^{8.0}_{6.7}$ & 2.0 & 22.569 & $0.41\pm^{0.45}_{0.31}$ & $15.53\pm^{.23}_{.33}$ &  
1.6(1) & 0.21 & \nodata{} & \nodata{}\\[5pt]
\hline
\ovii\ $z=0.4339$ (free)  & 119.7(87) & $486.0\pm^{5.2}_{5.4}$ & 2.5 & $30.975\pm^{0.009}_{0.007}$ & 
 $3.33\pm^{1.45}_{1.10}$  & $15.73\pm^{.13}_{.05}$ & 29.9(2) & $3.2 \times 10^{-7}$ &  $4\times  10^{-5}$ & <0.001\\[5pt]
            & & $476.0\pm^{4.5}_{4.6}$ & &  & & & &&\\[5pt]
\ovii\ $z=0.3551$ (free) & 117.6(90) & $814.6\pm^{8.6}_{7.2}$ & 2.0 & $29.283\pm^{0.020}_{0.008}$ & 
 $0.92\pm^{0.63}_{0.35}$ & $15.38\pm^{.22}_{.13}$ & 8.2(2) & $1.7 \times 10^{-2}$ & 0.88 & 0.46 \\[5pt]
                         &        & $780\pm^{7.5}_{5.5}$ & & & & & & &\\[5pt]
\hline
\hline
 \end{tabular}
 \end{table*}

\subsubsection{Results of the analysis}

The combination of the results shown in Table~\ref{tab:results} and \ref{tab:po} indicates that, of the eight redshift
systems with prior FUV absorption detected, there is only evidence for the detection of \ovii\ in
redshift system 1 at $z=0.1876$, with a null hypothesis probability of approximately 1\%. This null hypothesis probability,
if used in a standard normal distribution, corresponds to the probability to exceed $\pm 2.6$ standard deviations
\citep[see, e.g., Table~A.2 in][]{bonamente2022book}
and therefore it is equivalent to a 2.6--$\sigma$ level of significance for a possible detection. The significance of detection is obtained from the analysis of the $\Delta C$ statistic for the power--law plus \texttt{line} model, where the
conditions for the interpretation of the measured $\Delta C=6.6$ for a nested model component are satisfied. The \texttt{spline} plus \texttt{slab} model analysis (Table~\ref{tab:results}, $\Delta C=4.2$) cannot be immediately interpreted with a quantitative $p$--value, since the null--hypothesis value of the nested model component 
(column density $N=0$) is at the boundary of parameter space, as discussed in the previous section.

None of the other \ovii, \oviii\ or \neix\ lines appear to be present in the data. In fact, the second largest $\Delta C$ statistic
from Table~\ref{tab:results}  (for \oviii\ at redshift system 3) indicates that, upon reanalysis with the power--law plus \texttt{line}
model in Table~\ref{tab:po}, is clearly not significant (null hypothesis probability of 21\%).
It is also useful to point out that the possible \ovii\ detection has an estimated column density that 
is just above the $\log N (\text{cm}^{-2})=15.0$ limit associated with the systematic errors discussed in Sect.~\ref{sens.sec}.

A list of galaxies along the sightline towards \es\ and in a redshift range of $\pm 0.05$ of the \ovi\ redshift ($z=0.1876$) are reported in Table~\ref{tab:galaxies}. At that redshift, 1~arcmin separation corresponds to a transverse distance of $\sim 190$~kpc, for a standard WMAP9 $\Lambda$CDM cosmology \citep{wright2006,hinshaw2013}.  The smallest sight--line distance is 5 arcmin, i.e. 1 Mpc, or  $\sim 5 \times r_{200}$. The \texttt{EAGLE} simulations analysis by \cite{wijers2020} show
that \ovii\ column densities with impact parameters $1-3 \times r_{200}$ drop rapidly below  $\log N(\ovii)  = 15$ for galaxies of any mass. It is therefore clear that even the galaxy at the closest impact parameter is not sufficiently close to the sightline to be able to cause the tentative \ovii\ absorption line.

\begin{table*}
\centering
\caption{Galaxies along the sightline towards \es\  at redshifts $z=0.182-0.192$, from the \emph{NASA Extragalactic Database}.}
\label{tab:galaxies}
 \renewcommand{\arraystretch}{1}
 \hspace*{-2.5cm}
 \begin{tabular}{c c c c c c c}
 \hline
No. & Object Name & RA (degrees) & DEC (degrees) & Redshift & Magnitude and Filter & Separation  (arcmin)  \\
\hline
1 & WISEA J155552.00+111536.2 & 238.96668 & 11.26004 & 0.18939 & 19.1g & 4.74 \\  
2 & WISEA J155556.00+111558.8 & 238.98332 & 11.26633 & 0.1875  & 18.9g & 5.57 \\  
3 & WISEA J155517.24+111307.3 & 238.82160 & 11.21872 & 0.18572 & 20.7g & 6.57  \\  
4 & WISEA J155521.73+110557.5 & 238.84056 & 11.09932 & 0.18823 & 19.7g & 7.55 \\  
5 & WISEA J155507.95+111208.2 & 238.78314 & 11.20226 & 0.19089 & 20.0g & 8.64 \\  
6 & 2MASS J15551843+1104530   & 238.82681 & 11.08144 & 0.18777 & 19.4g & 8.89 \\  
7 & WISEA J155525.65+111955.5 & 238.85674 & 11.33199 & 0.18459 & 19.7g & 9.53 \\  
\hline
\end{tabular}
\end{table*}

\begin{table}
    \centering
    \begin{tabular}{l|ll}
    \hline
    \hline
    Parameter & \multicolumn{2}{c}{Value} \\
    & OVII & OVIII \\
    \hline
    \hline
        \multicolumn{3}{c}{RGS1 Power-laws (continua)}\\
        \hline
        $\alpha_1$ & 1.32$^{+ 0.21}_{-0.21}$&  4.82$^{+0.19}_{-0.17}$\\
        Norm$_1$ &  1435$^{+235}_{-204}$& 145$^{+16.1}_{-16.0}$\\
        Range (\AA)& 25-27 & 22-24 \\
        \hline
        $\alpha_2$ & 1.48$^{+0.42}_{-0.41}$& 1.51$^{+0.19}_{-0.18}$\\
        Norm$_2$ &  1032$^{+276}_{-222}$& 1011$^{+92.0}_{-86.3}$\\
        Range (\AA)&21.5-23.1 &19.1-20.7\\
        \hline
        $\alpha_3$ & 1.0 (Fixed)& 2.80$^{+0.61}_{-0.64}$\\
        Norm$_3$ & 1868$^{+17}_{-17}$& 464$^{+299}_{-171}$\\
        Range (\AA)&29.5-30.5 &26-27\\
        \hline
        \multicolumn{3}{c}{RGS2 Power-laws (continua)}\\
        \hline
        $\alpha_1$ & 1.69$^{+0.19}_{-0.19}$&--\\
        Norm$_1$ & 1061$^{+157}_{-138}$&--\\
        Range (\AA)& 25-27 & 22-24 \\
        \hline
        $\alpha_2$ &--& 2.11$^{+0.41}_{-0.44}$\\
        Norm$_2$ &--& 742$^{+164}_{-124}$\\
        Range (\AA)&21.5-22.8 &19.1-20.7\\
        \hline
        $\alpha_3$ & 0.049$^{+0.74}_{-1.14}$& 1.80$^{+0.64}_{-0.64}$\\
        Norm$_3$ & 4241$^{+6850}_{-2012}$& 970$^{+598}_{-373}$\\
        Range (\AA)&29.5-30.5 &26-27\\
        \hline
        \multicolumn{3}{c}{Stacked \ovii\ Absorption Line Model}\\
        \hline
        Redshifts & 1, 3, 4, 5, 6, 8 \\
        Line $\tau_0$ (linked) &  0.27$^{+0.13}_{-0.12}$\\
                            & (or $\log N_{\text{OVII}} (\text{cm}^{-2})=14.94\pm^{0.18}_{0.22}$) \\
        \hline
        DOF & 325\\
        $C$--stat (expect.)& 380.01  (337.11 $^{+}_{-}$25.97)\\
        $\Delta C$  &   5.02\\
        $p$--value & 0.025\\
        \hline
        \multicolumn{3}{c}{Stacked \oviii\ Absorption Line Model}\\
        \hline
        Redshifts & 1, 3, 4, 5, 6, 7\\
        Line $\tau_0$ (linked) & 0.15$^{+0.082}_{-0.13}$ \\
                              & (or $\log N_{\text{OVIII}} (\text{cm}^{-2})=15.07\pm^{0.17}_{0.87}$) \\
        \hline
        DOF & 272\\
        $C$--stat (expect.) &  323.04 (285.07 $^{+}_{-}$23.88)\\
        $\Delta C$  &  1.57\\
        $p$--value & 0.21\\
        \hline
        \hline

    \end{tabular}
    \caption{Spectral fit for the stacked \ovii\ and \oviii\ WHIM lines. In order to include all possible lines, the RGS data were fit
    in three narrow--band intervals (i.e., 25-27\AA, 21.5--23.1~\AA\ and 29.5--30.5~\AA\ for \ovii) with \texttt{line} models at 
    the expected wavelengths for the redshift systems under consideration (e.g., redshift systems 1,3,4,5,6,7 and 8 for \ovii). The normalization
    and optical depth at line center ($\tau_0$) for the \texttt{line} models were linked among all the lines. Power--law index $\alpha_3$ for the \ovii\ fits was fixed, due to calibration uncertainties in that band.}
    \label{tab:stack}
\end{table}

\begin{table*}
 \caption{ Results of the power--law plus \texttt{line} model for the \chandra\ LETG data.
 The $\Delta C$ statistic refers to zeroing out the nested \texttt{line} component. The rightmost 
 column reports the  $\Delta C$ statistic corresponding to the comparison between the best--fit model, and a model where the \texttt{line} component was fixed at the \xmm\ RGS best--fit values from Table~\ref{tab:po}, instead of the best--fit from \chandra\ data (this table).}
 \label{tab:poChandra}
 \begin{tabular}{c|lllllll|ll}
 \hline
 \hline
 Target line & \cmin& \multicolumn{2}{c}{power--law component} & \multicolumn{3}{c}{line component}& & \multicolumn{2}{c}{RGS best-fit} \\
             & (d.o.f.)       & \text{norm.} & \text{index} & $\lambda$ (\AA) & $\tau_0$ & $\log N (\text{cm}^{-2})$ & $\Delta C$ & $\Delta C$ & d.o.f. \\
 \hline 
\ovii\ $z=0.1876$ (\# 1) & 80.66(77) & $1510\pm^{31}_{31}$ & 2.0 & 25.6545  & $0.32\pm^{0.60}_{0.46}$ & $15.03\pm^{0.34}_{}$ & 0.44 & 0.2  & 1\\
\oviii\ $z=0.1898$ (\#3) & 28.82(30) & $1369\pm^{35}_{35}$ & 2.0 & 22.5693 & $1.37\pm^{2.2}_{1.0}$ & $15.87\pm^{0.22}_{0.45}$ & 2.09 & \nodata & \nodata\\[5pt]
\hline
\ovii\ $z=0.4339$ (free) & 92.45(76) & $876\pm^{21}_{20}$ & 2.5 & 31.0138 & $0.51\pm^{0.7}_{1.0}$ & $15.23\pm^{0.31}_{}$ & 0.77 & 10.9 & 2\\
\ovii\ $z=0.3551$ (free) & 97.26(76) & $1426\pm^{33}_{32}$ & 2.0 & 29.3492 & $2.06\pm^{1.1}_{1.1}$ & $15.70\pm^{0.11}_{0.24}$ & 6.58 & 5.3 & 2\\[5pt]
\hline
\hline
 \end{tabular}
 \end{table*}

\begin{table*}
\caption{Selected parameters of the \texttt{spline} model used in the fits of Table~\ref{tab:results}. Parameter $y001$ refers to the \texttt{spline} component at the lowest wavelength considered $\lambda=13$\AA, and $y014$ for that at $\lambda=26$\AA\ .}
\label{tab:spline}
 \renewcommand{\arraystretch}{1.2}
 \begin{tabular}{l c c c}
 \hline\hline
        \# & $z$  & y001 & y014\\
 \hline
1 & $0.1876$ & $ 990^{+20}_{-21}$  & $5178^{+50}_{-45}$ \\
3 & $0.1898$ & $ 993^{+20}_{-24}$  & $5178^{+49}_{-42}$ \\
4 & $0.3950$ & $ 977^{+19}_{-20}$  & $5251^{+42}_{-43}$ \\
5 & $0.0347$ & $ 970^{+193}_{-135}$& $5182^{+372}_{-401}$ \\
6 & $0.0427$ & $ 995^{+20}_{-23}$  & $5149^{+51}_{-88}$ \\
7 & $0.0636$ & $ 966^{+193}_{-134}$& $5168^{+388}_{-384}$ \\
8 & $0.2186$ & $ 994^{+21}_{-24}$  & $5143^{+79}_{-64}$ \\
\hline
\multicolumn{4}{c}{Redshifts from \protect\cite{nicastro2018}}\\
9 & $0.4336^{+0.0005}_{-0.0003}$  & $ 974^{+28}_{-23}$  & $ 5224^{+58}_{-51}$\\
10 & $0.3557^{+0.0006}_{-0.0006}$ & $ 995^{+19}_{-23}$  & $ 5164^{+43}_{-45}$\\

\hline
\hline
\end{tabular}
\end{table*}

\subsubsection{Additional Stacking of data}
\label{sec:stack}
Given the limited evidence for the detection of individual absorption lines from a given
redshift system, we experimented with stacking the data in order to improve the statistical sensitivity of the data towards the more numerous absorbers expected at the lower column densities. 
We estimated the statistical sensitivity of the data in terms of column density as follows. 
We applied the same
\texttt{slab} model used for Table~\ref{tab:results} to a region near 26~\AA\ where both RGS1 and RGS2 data are present, and away from detector artifacts. We fit the spectrum and use the $\Delta C=2.7$ criterion
to identify an upper limit of $\log N_\mathrm{OVII} (\text{cm}^{-2}) \simeq 15.3$, which corresponds to a 90\% upper limit
to the non--detection of \ovii. 
When combining or `stacking' $n$ segments of the spectrum at pre--specified 
wavelengths, the sensitivity of the stacked spectrum is expected to be improved
(i.e., the column density that can be probed is reduced) by a factor of
approximately $\sqrt{n}$, based on the reduction of the standard deviation
of the average signal by the same factor. For $n=7$,  the stacked data are therefore
expected to reach a sensitivity of approximately $\log N_\mathrm{OVII} (\text{cm}^{-2}) \simeq 14.9$. This is the level of \ovii\ column density we expect to probe with the stacking of the spectra.

We stacked a $\pm 0.5$~ \AA\ portion of the spectra, centered at the target 
redshifted absorption line, for all redshift FUV priors for the \ovii\ and \oviii\ ions. 
To accomplish this, we performed a different 
spectral fit, compared to the one that resulted in Table~\ref{tab:results}. This fit consists of using
three narrower portions of the RGS1 and RGS2 spectra, each successfully fit with a simple
power law, in such a way that the wavelengths of the FUV \ovii\ and \oviii\ absorption line models (\texttt{line} in SPEX) 
were all included in these regions. 
For each of these two fits, the \texttt{line} models were linked to one another, so that there is only one
free parameter ($\tau_0$) describing the combined optical depth of the lines at the line center.
This method was preferred to the one performed, for example, in the stacking analysis of \cite{kovacs2019}, where
the spectra are first shifted, and then fit to a model that includes an absorption line. The reason for this choice was
primarily due to the fact that fitting in observed (i.e., redshifted) wavelength space preserves more accurately the 
statistical properties of the data, without being affected by the additional processing that results from the
blue--shifting to the rest--frame wavelength.
Details of these additional spectral fits are reported in
Table~\ref{tab:stack}.  We decided to not stack the signal at the \neix\ wavelengths in order to focus on the high--temperature oxygen ions alone, given that the only possible detection is from \ovii.

The results of these stacked analyses is that there is marginal evidence for the detection of
possible \ovii\ absorption at the combined FUV--prior wavelengths, 
with a null hypothesis probability of 2.5\%, or approximately a 2.2--$\sigma$ significance. 
The stacking permits to reach a lower average column density of $\log N_{\mathrm{OVII}} (\text{cm}^{-2}) =14.9\pm0.2$, which is consistent with the simple $\sqrt{n}$ improvement in the sensitivity discussed at the beginning of the section.
This column density from the staked data is consistent with 
 \ovii\ being present only at redshift system 1. 
While the expected number of absobers per redshift increases by the stacking procedure, the
relatively small redshift path renders the detection probabilities small. 
We will investigate this issue in more detail in a future work with a larger redshift path.
%
The stacked \xmm\ data do not reveal any evidence for \oviii\ absorption at the combined FUV--prior wavelengths.

In addition to the spectral fits, and for the purpose of
showing a stacked blue--shifted spectrum in the rest frame, we also combined the relevant regions of the spectrum in Figure~\ref{fig:oviistack}. 
For this figure, $\pm 0.5$~ \AA\ portions of the spectra were blue--shifted
to the rest--frame wavelength of interest, and the data and models added together. In addition,
the data are represented as their ratio with respect to the best--fit model, in such as way that the orange points are expected to scatter around the 1.0 value of the ratio. To further illustrate the effect of the best--fit
\texttt{line} model, this component was zeroed out to calculate the ratio of data--to--model now
represented by the blue data points. The  blue data points show
that, for the \ovii\ stacking, there is a preponderance of points below the 1.0 value around the 
rest--frame wavelength center of 21.6~\AA, indicating a preference for
absorption at those wavelengths (although,as discussed earlier, only with limited significance).
The black line further reports the difference between the full model and the model with the line component 
zeroed out, to illustrate the depth of the best--fit absorption line model for the stacked \ovii\ and \oviii\ 
lines with FUV priors. This figure is for illustration purposes only, since the fitting was performed
in the proper observed wavelength frame, as reported in Table~\ref{tab:stack}.

\subsection{Analysis of the two serendipitous \protect\cite{nicastro2018} lines}
\label{sec:nicastro}
In this section we address the \cite{nicastro2018} report of the serendipituous detection of
putative strong He--$\alpha$ \ovii\ absorption at two redshifts with previously unreported FUV absorption lines. 
For this purpose, we use the same model as for the FUV priors, with an initial value for the
redshift at the best--fit values reported by the previous study (respectively, $z=0.4339$
and $z=0.3551$), but with the redshift of the \texttt{slab} model free to adjust itself to the
best--fit value. Results of these fits are reported at the bottom of Table~\ref{tab:results}. 
We also used the simplified power--law plus line model described in Sect.~\ref{sec:FUVpriors}
for the two possible \ovii\ lines,
so that we can more readily use the corresponding $\Delta C$ values for hypothesis testing.
The results of these additional fits are also reported at the bottom of Table~\ref{tab:po}, where we left
the normalizations of the RGS1 and RGS2 data uncoupled to provide more flexibility in the fit.
The relevant portions of the \xmm\ spectra are shown in Fig.~\ref{fig:nicastro}, with the
red dotted line indicating the redshift of the lines that was identified by the \cite{nicastro2018} analysis.

As discussed in Sect.~\ref{sec:redshift}, there are uncertainties in our knowledge of the redshift
to \es, with a most likely redshift of $z\simeq0.433$. If this is correct, then the serendipitous absorber at
$z=0.4339$ would be intrinsic to the source, and unlikely to be associated with the WHIM. 
Nonetheless, we proceed with the
study of both of the reported \cite{nicastro2018} absorbers, \emph{assuming} their WHIM origin and further
using their $z \leq 0.48$ limit on the source's redshift.

\subsubsection{Statistical analysis of the serendipitous detections}
\label{sec:redshiftTrials}
In order to evaluate the statistical significance detection of a serendipitous line, i.e.,
of a line that did not have a predetermined wavelength, it is necessary to account for the
number of  `redshift trials', or independent opportunities to detect such feature, as originally
proposed by \cite{kaastra2006}.
A serendipitous or blind search is such that there are multiple opportunities to interpret the
deepest fluctuation in the data as a possible absorption line. 

\begin{figure}
    \centering
    \includegraphics[width=3.4in]{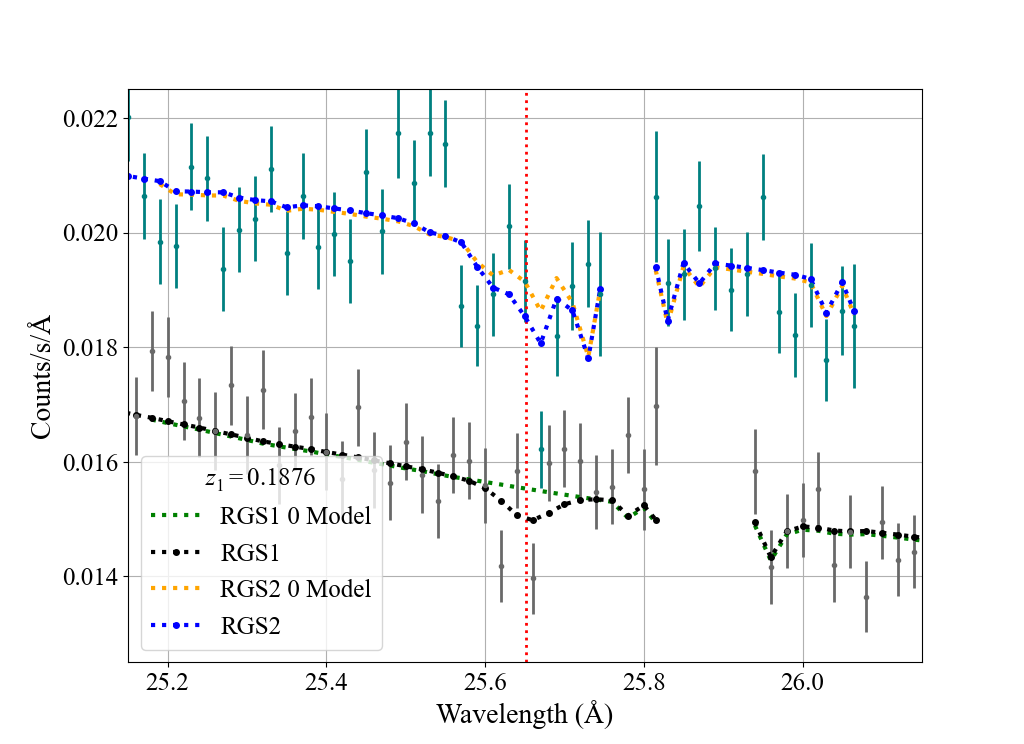} 
    \includegraphics[width=3.4in]{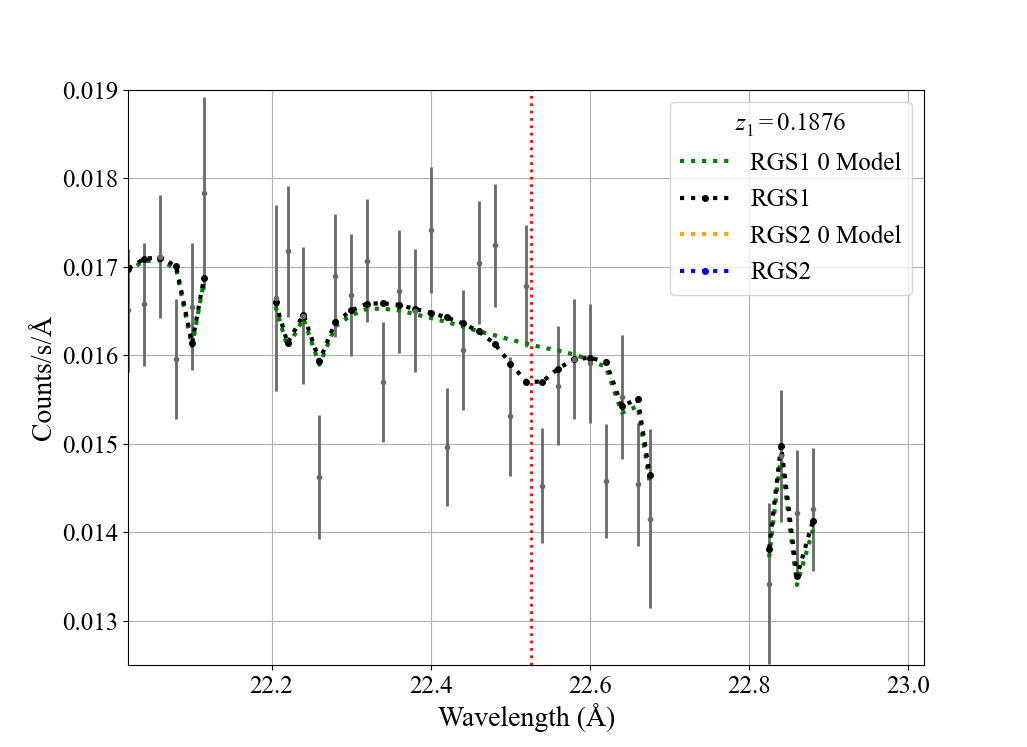} 
    \includegraphics[width=3.4in]{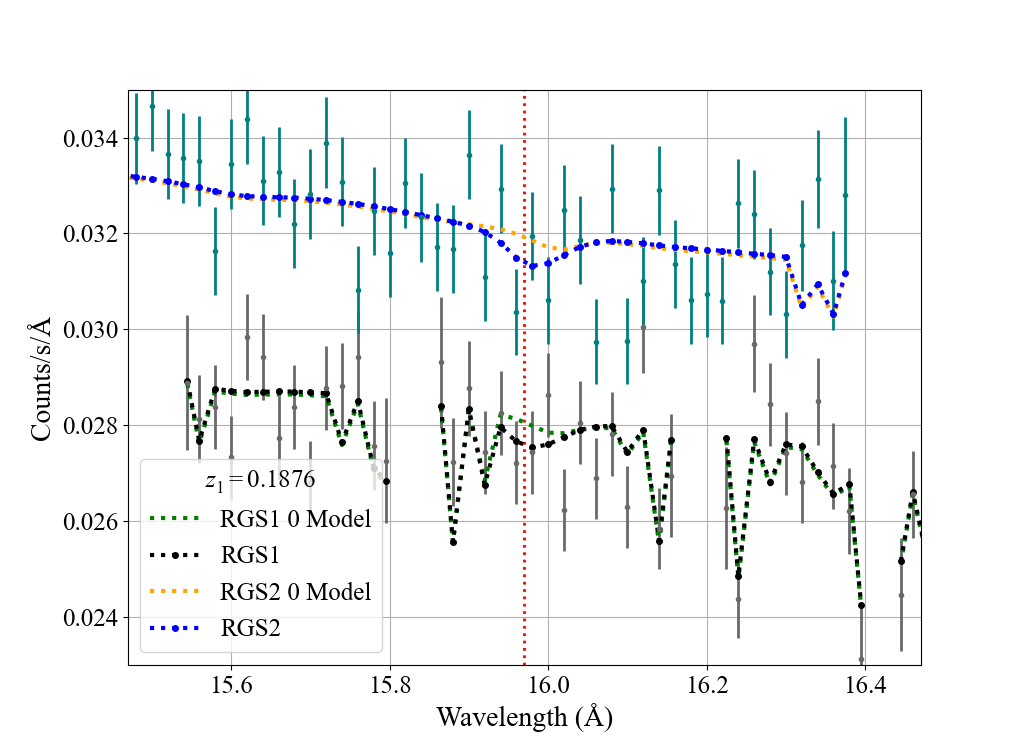}
    \caption{From top to bottom: Portion of the \xmm\ spectra near the 
    O~VII, O~VIII and Ne~IX at $z=0.1876$. In the top and bottom panels,
    the RGS2 data (blue) were shifted by a factor of 1.2 for clarity. Vertical lines mark the wavelengths of the 
    redshifted lines. 
    }
    \label{fig:z0.1876}
    \end{figure}

A general method to evaluate the probability
of occurrence of such fluctuations in a blind search is that of performing a numerical simulation
that includes all the relevant parameters of the search, such as the wavelength range spanned, the 
shape of the line--spread function (LSF), and the resolution of the instrument. Accordingly, we performed
a simulation that uses the \xmm\ LSF model as parameterized by \cite{kaastra2006}, with line center 
that can vary between $\lambda_1=23.5$~\AA\ and $\lambda_2=32$~\AA. The first wavelength 
enforces $z \geq 0$ and avoids the 21.6-23.5~\AA\ range possibly contaminated by several \oi--\ovi\ Galactic lines,
and the second wavelength enforces $z \leq 0.48$, which is the minimum possible redshift for the
source reported by \cite{nicastro2018} although, as remarked in Sect.~\ref{sec:redshift}, there is now
evidence that the redshift of the source may be $z\simeq0.433$. In addition, and somewhat conservatively, a 25\% reduction of this range
is assessed in an attempt to correct for several small regions that are unavailable 
due to detector efficiency and calibration issues (reported, for example, as blank portions in the
spectra of Figure~\ref{fig:z0.1876}), resulting in a search over a wavelength range of size $\Delta  \lambda=6.4$~\AA.

To emulate the blind search method in our numerical simulation,  we stepped the possible 
absorption--line LSF model by a wavelength interval that is smaller than the 20~m\AA\ data bin (i.e., by 10~m\AA), and identified the
distribution of the $\Delta C$ statistic associated with the \emph{strongest negative fluctuation} during this search, by comparing
two fits obtained, respectively, with and without the line component. 
Since our data are in the large--count limit in each bin,
with at least 400 counts per bin, the $\Delta C$ statistic was approximated by the corresponding
$\Delta \chi^2$ statistic, assuming Gaussian distribution for the counts in each bin. It is important to point out that,
as also suggested by \cite{kaastra2006}, the details of the continuum model
are not important, since the fit can be performed on the standardized deviations from such best--fit model, where each 
bin follows a standard Gaussian distribution, under the hypothesis that the data follow the model.
The distribution of this maximum $\Delta \chi^2$ of the search, which approximates the sought--after distribution of
the maximum $\Delta C$, is reported in Figure~\ref{fig:DeltaCBlind}.

The distribution of Figure~\ref{fig:DeltaCBlind} can be immediately used to identify critical values of the
 $\Delta C$ statistic, and thus determine whether the measured statistics for a given line is significant or not.
 The $p$--values associated with this statistic will be referred to as the `corrected $P$ value' in Table~\ref{tab:po}.
 According to this analysis, the $z=0.4339$ serendipitous line has a corrected $P$ value that remains comfortably in 
 excess of the 99.9\% confidence level, given that the $p=0.999$ critical value of the distribution is estimated at $\sim 20.0$.
 On the other hand, the putative $z=0.3551$ has a corrected $P$ value of $\sim$0.46, or a 46\% probability that
 the fluctuation is consistent with the noise level. We therefore conclude that
 this line is very likely a statistical fluctuation in the data.
 The method is less accurate in the determination of the exact $p$--value of the measured statistic, especially for lines with a large value of $\Delta C$ such as the first of the two \cite{nicastro2018} lines, since it is difficult to
 simulate the tail of the distribution accurately.  

An alternative and approximate method to determine the $p$--value of the detection statistics 
is based on the use of 
the binomial distribution, and it is described in detail in \cite{bonamente2019}. According to this approximate method,
the number $N$ of independent opportunities to detect an absorption line is estimated
as 
\[ N =\dfrac{\Delta \lambda}{\sigma_{\lambda}}, \] 
where the numerator is the effective wavelength range of the search, accounting for unavailable portions
of the spectrum, and the denominator is the resolution of the instrument, interpreted as a characteristic
wavelength range required to detect a line or to separate one line from a neighboring one.
The first number can be be estimated taking into account the parameters of the search, as 
described earlier, with $\Delta \lambda \simeq 6.4$~\AA.
The second number can be estimated to be of order $\sigma_{\lambda} \simeq 50$~mA, which is the approximate
resolution of the RGS spectrometers; this method is therefore approximate in that it does not
account explicitly for the shape of the LSF, but just the approximate resolution. With these numbers, it is possible to estimate 
that approximately $N \simeq 127$ independent opportunities were available to detect 
a serendipitous \ovii\ WHIM absorption line in the spectrum of \es.

The next step is to determine the statistical significance of the detection, 
indicated 
by the null hypothesis probability $P$, and its relationship with the 
single--trial probability indicated with the usual lower--case $p$. The simplest
approximation discussed in \cite{bonamente2019} and also by \cite{nicastro2013} is
\[ P \simeq p N \]
assuming $p N \ll 1$, which is applicable to this case.
The corresponding corrected $P$ values are reported  in the last column of Table~\ref{tab:po}.
The redshift--trials--corrected null hypothesis probability for the 
first line is $P=4 \times 10^{-5}$, meaning that there is just a 0.004\% probability that this is 
a chance fluctuation. This is consistent with the results based on the simulation of the $\Delta C$ statistic
of Fig.~\ref{fig:DeltaCBlind}.
 It is therefore possible to
conclude that there is strong evidence for the serendipitous detection of a 
genuine absorption line feature in the spectrum of \es, which corresponds to
\ovii\ He-$\alpha$ at $z=0.4339$. When this null hypothesis probability $P$, inclusive
of the number of redshift trials,
is reported in terms of a two--sided hypothesis testing with the Gaussian distribution,
it corresponds to a 4.1-$\sigma$ level detection. This significance of detection is similar to that reported by \cite{nicastro2018}, 
well in excess of the 3--$\sigma$ threshold even when
accounting for all possible redshift trials.

For the other serendipitous \ovii\ line at $z=0.3551$ reported by \cite{nicastro2018},
it is already evident from the larger $p$ value at a fixed redshift (well below the 3--$\sigma$
threshold) and the corrected $P$ value based on the simulation of the $\Delta C$ statistic
that there is no significant fluctuation. This is also consistent with the re--analysis
by \cite{nicastro2018b}, where the author updated the initial significance of detection for this
possible line and indicated that there is no significant evidence for an absorption line at that
redshift. It is nonetheless instructive to evaluate 
the overall probability inclusive of redshift trials also for the second putative line, based on the approximate 
binomial distribution model. 
In this case, the $P$ value  cannot be calculated according to the simple $pN$ approximation, since this 
product is a large number. Instead, one needs to follow the binomial probabilities (see Eq.~5 
of \citealt{bonamente2019}), which consists of evaluating the
sum
\[ 
P= \sum_{i=1}^N \binom{N}{i} p^i (1-p)^{N-i}
\]
where $p$ is the standard single--trial $p$--value, in this application $p=0.017$. This
calculation results in a redshift--trial corrected probability of $P=0.88$, confirming that this is most likely a random fluctuation, and similar to the $P=0.46$ value obtained according to the distribution of Fig.~\ref{fig:DeltaCBlind}.

\subsubsection{Other considerations for these serendipitous features}

Having established that the first of the two \cite{nicastro2018} features
is consistent with \ovii\ He--$\alpha$ absorption at $z=0.4339$, it is necessary to discuss
its identification. In the absence of a more significant FUV counterpart at the same redshift,
the identification with \ovii\ He--$\alpha$ should be regarded as indeed plausible, but only tentative. 
The accompanying He--$\beta$ would fall near a wavelength of 26.71~\AA, but its lower oscillator strength (see Table~\ref{tab:Lines})
makes the line less readily detectable. \cite{nicastro2018}, in fact, does not report a statistically significant detection
at those wavelengths. There is also the possibility of a misidentification with other Galactic 
or intervening lines. One such possibility is Galactic N~II K--$\alpha$ absorption,
which is expected to be present at the same wavelengths as the detected feature, and may in fact be responsible for a portion of the
detected absorption \citep[as already discussed in][]{nicastro2018b}.  

Moreover, \cite{johnson2019} showed that \es\ is likely a member of a group of galaxies
located near $z=0.433$, although its redshift could not be measured due to the lack of emission lines in
its optical spectrum. The possible location of the source at the same redshift
as the putative \ovii\ absorption lines results in the possibility that 
the feature could be intrinsic to the source itself, also also discussed in \cite{nicastro2018},
or that the absorption is associated with the local intra--group medium, rather than the truly
inter--galactic WHIM.
It goes beyond the scope of this paper to investigate in more detail the identification of this serendipitous feature.
In the following, we will assume that it is entirely due to \ovii\ He--$\alpha$ absorption 
from the WHIM at $z=0.4339$, as in
the original \cite{nicastro2018} paper, and simply caution the reader that this identification is not confirmed,
and it needs to be regarded as tentative.

\begin{figure*}
\includegraphics[width=3.4in]{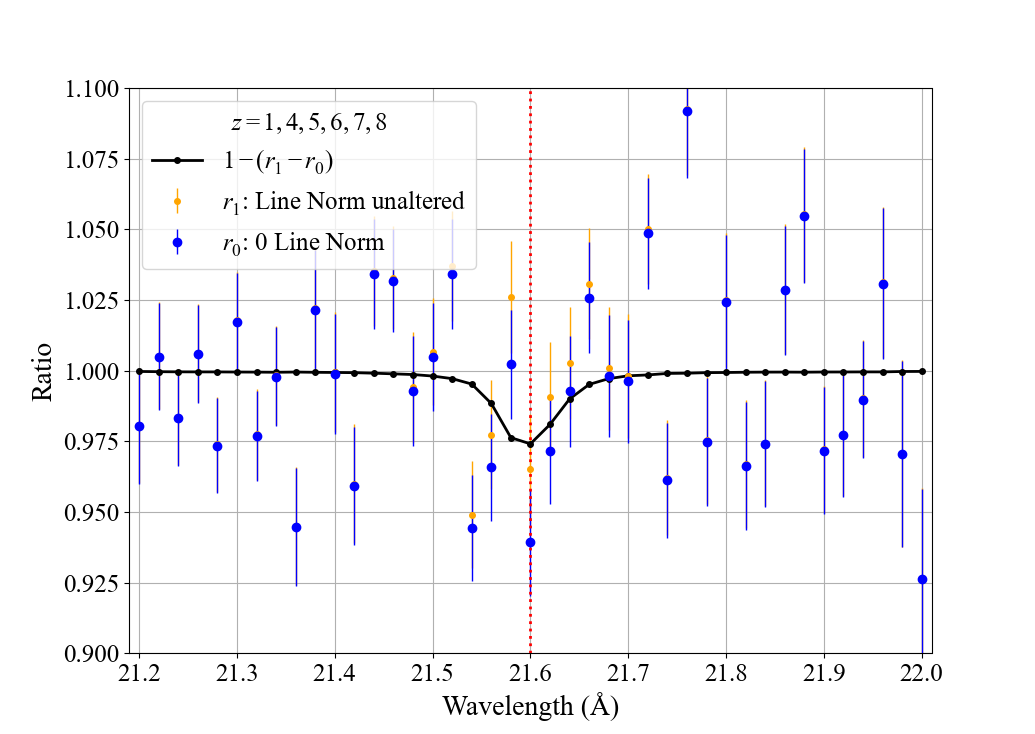}
\includegraphics[width=3.4in]{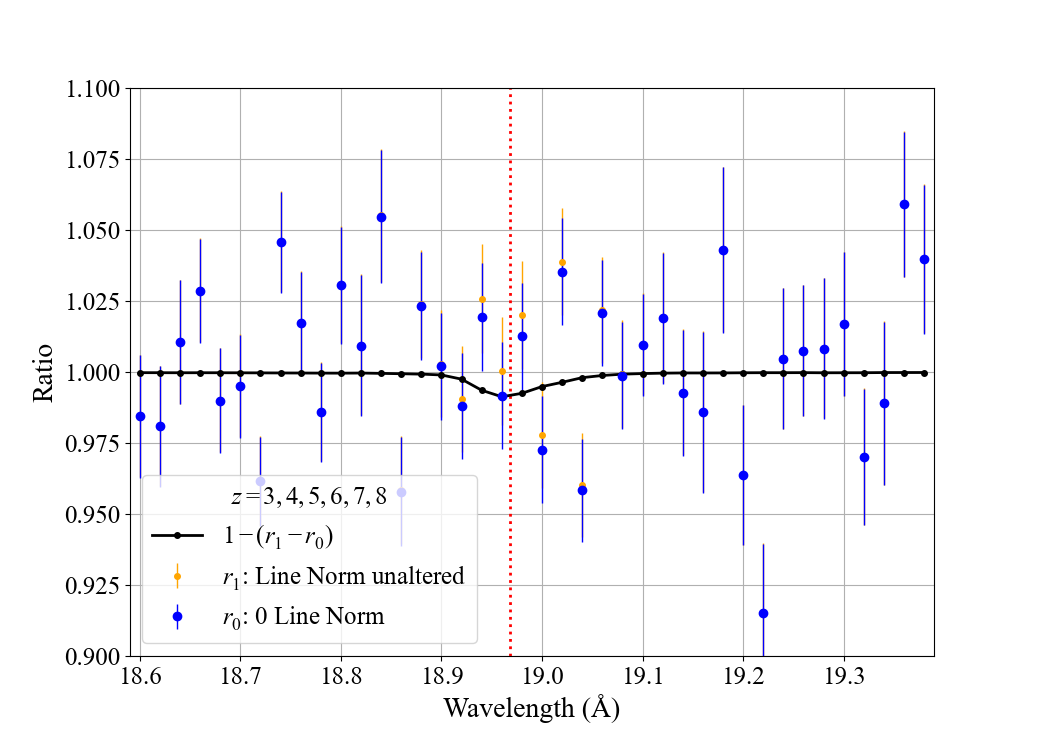}
\caption{(Left) Stacked spectrum for all \ovii\ lines with FUV priors. The orange data points are the ratio of the stacked data to the model, and the blue data points are the ratio when the OVII line model is zeroed out, to highlight any deficit of counts near the
expected line wavelength. In black is the difference between these two ratios. (Right) Same spectrum, but for
\oviii\ lines with FUV priors. The vertical lines mark the rest wavelength of the lines.}
\label{fig:oviistack}
\label{fig:oviiistack}
\end{figure*}

    \begin{figure*}
    \includegraphics[width=3.4in]{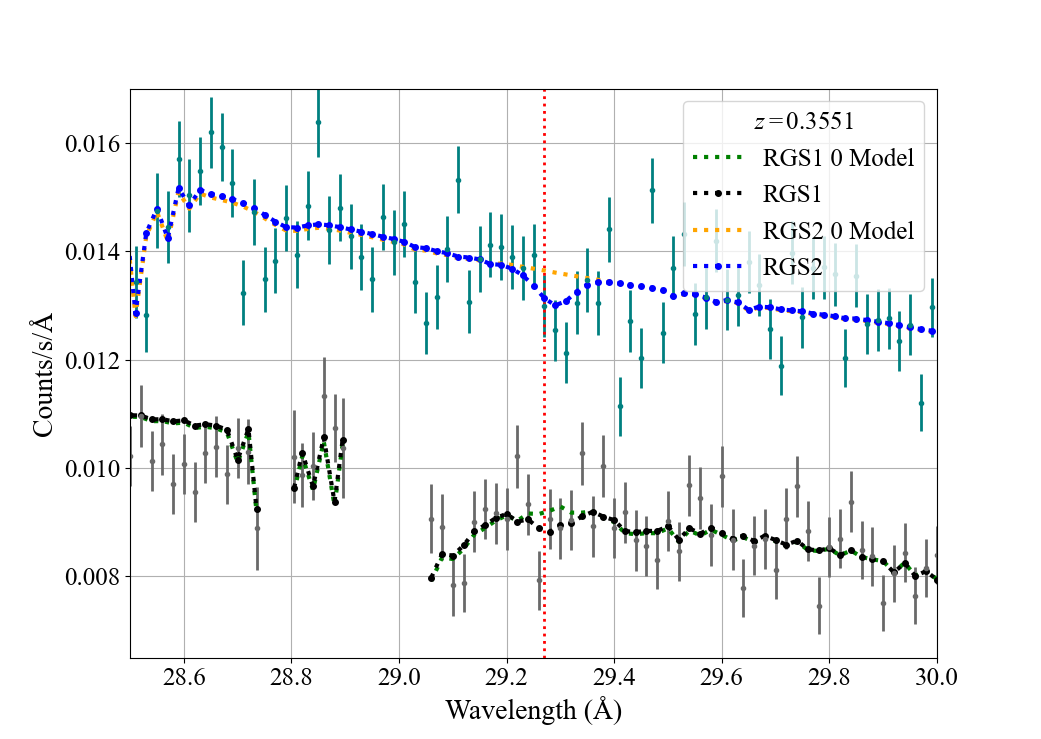}
    \includegraphics[width=3.4in]{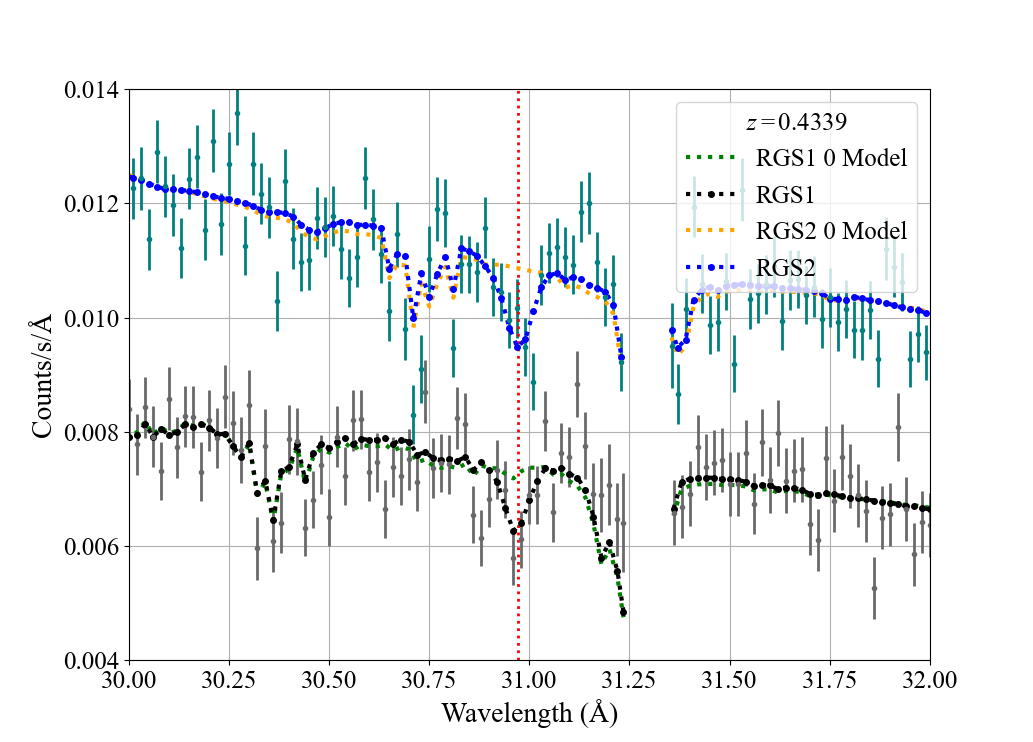}
    \caption{(Left) O~VII at $z=0.355$ and (Right:) O~VII at $z=0.4339$, at the 
    redshifts identified in the blind search of \protect\cite{nicastro2018}. Gray/black points represent RGS1, and teal/blue represent RGS2 (shifted for clarity by a factor of 1.35). Vertical lines are the
    wavelengths of the \protect\cite{nicastro2018} detections. }
    \label{fig:nicastro}
\end{figure*}

\begin{figure}
    \centering
    \includegraphics[width=3.4in]{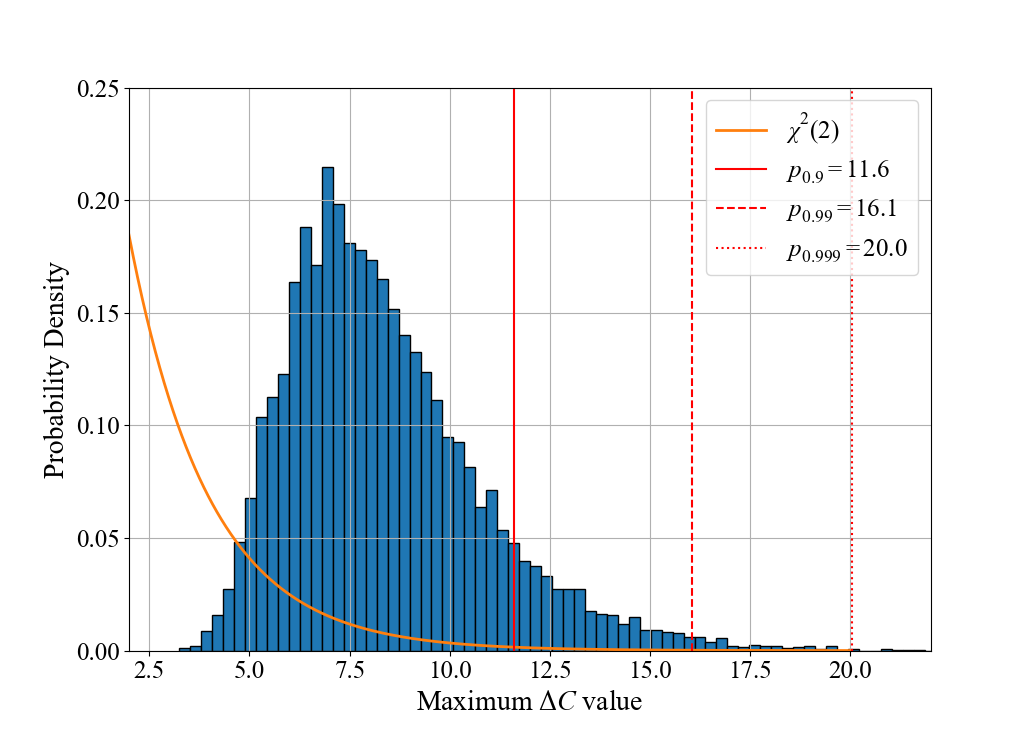}
    \caption{Sampling distribution of the $\Delta C$ statistic for the blind search of an absorption lines,
    under the null hypothesis that there are no absorption lines, following the method of \protect\cite{kaastra2006}. 
    For comparison it is also
    plotted the $\chi^2(2)$ distribution, which applies to the detection of an absorption line model with two
    free parameters, but without accounting for redshift trials.}
    \label{fig:DeltaCBlind}
\end{figure}

\subsection{Comparison of \xmm\ results with the \chandra\ data}
\label{sec:chandra}
\begin{figure*}
\centering
\vspace{3cm}
\includegraphics[width=3.2in]{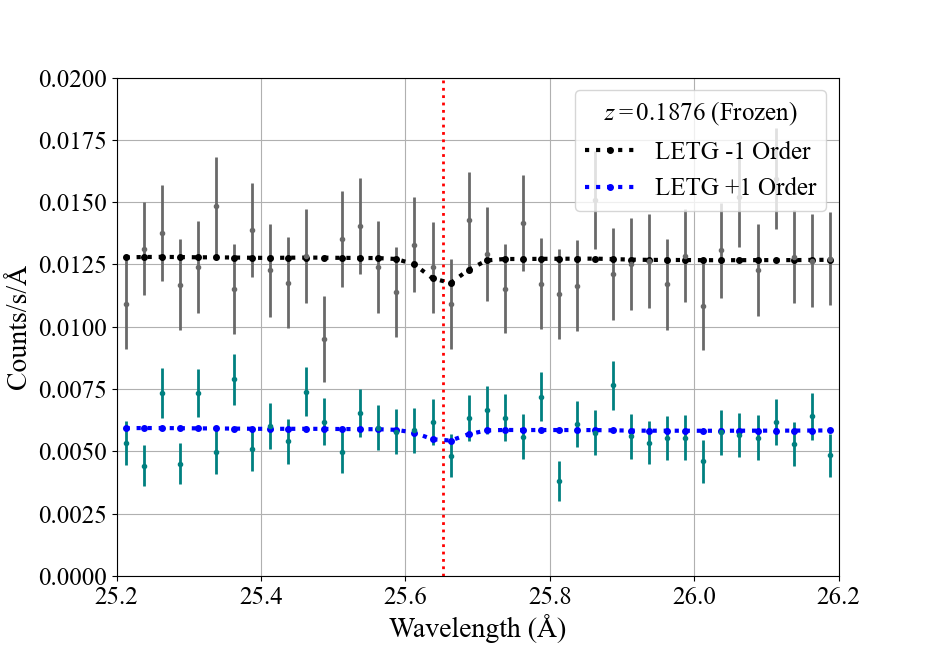}
\includegraphics[width=3.2in]{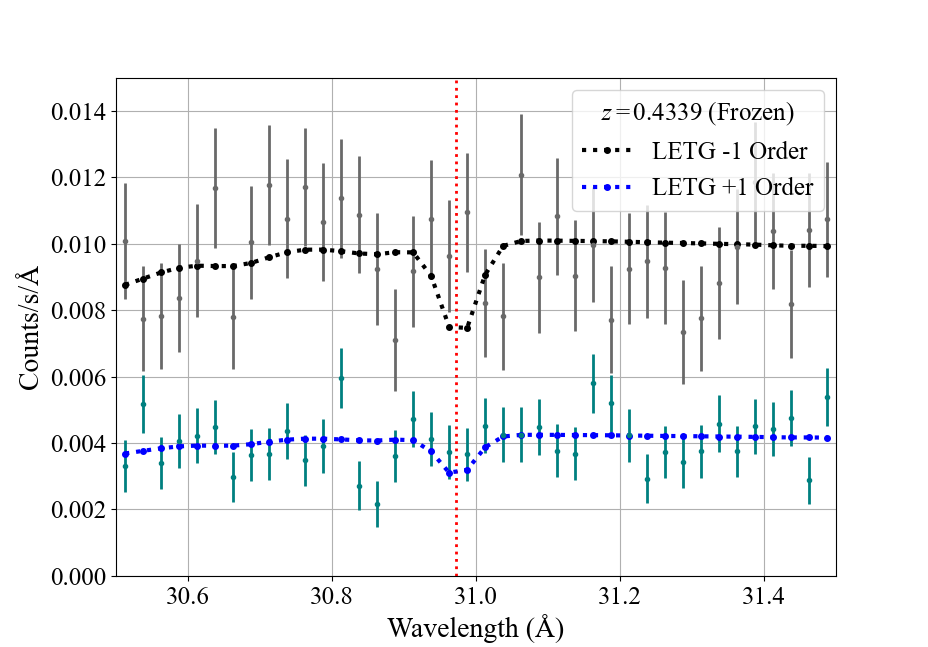}
\includegraphics[width=3.2in]{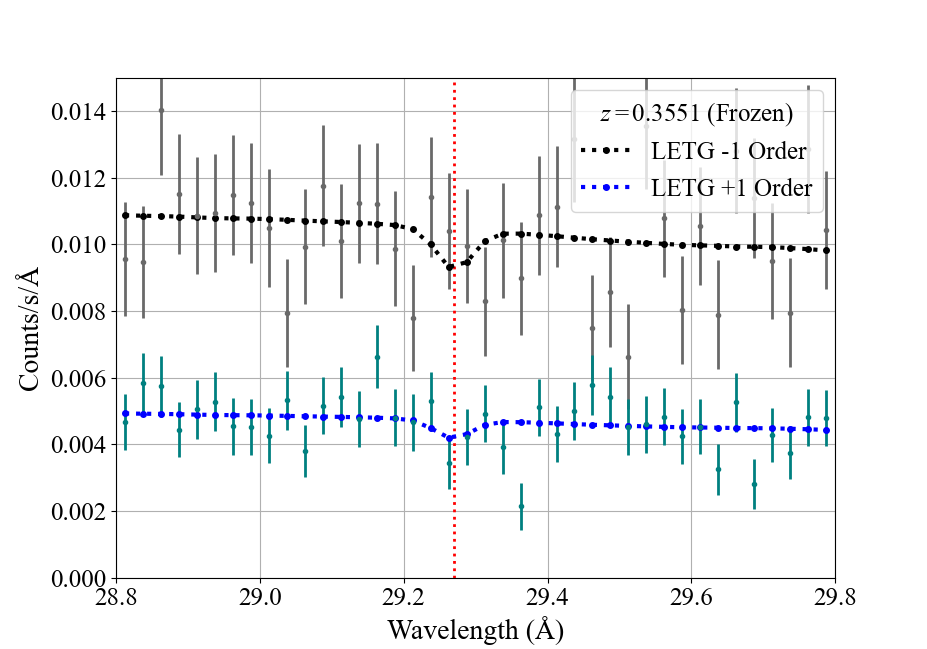}
\caption{ \chandra\ HRC LETG spectra in 1~\AA\ intervals around the three
wavelengths of interest. The -1 order spectra were rescaled by a factor of two, for clarity.}
\label{fig:chandra}
\end{figure*}

Despite the lower resolution, it is nonetheless useful to
determine whether the \chandra\ spectra are consistent with the
results from the \xmm\ spectra.
The \chandra\ HRC spectra near the wavelengths of the putative $z=0.1876$
\ovii\ absorption, and for the two \cite{nicastro2018} lines,
are shown in Fig.~\ref{fig:chandra}. For each of the spectra, we fit a simple
power--law model with a \texttt{line} component, same as in the \xmm\ fits of
Table~\ref{tab:po}. To compare the results between the two instrument, in one
of the fits the parameter(s) of the line--model component were left free, and in the
second fit they were fixed at the best--fit parameters from the \xmm\ regressions.
Table~\ref{tab:poChandra} shows the results for these fits, including the 
$\Delta C$ statistic for the comparison 
between the free--parameter fits, and the fits with the line component
fixed at the \xmm\ value. 

The best--fit model for the  \texttt{line} component shows that none
of the three lines is detected significantly, as can be seen from Fig.~\ref{fig:chandra}. In particular, the putative serendipitous \ovii\
$z=0.4339$ line is not detected at all in \chandra\ at the wavelength
and optical depth expected based on the \xmm\ spectra, despite its strong
\xmm\ detection. To determine the agreement between the two spectra
for that line, we determine that the fixed \xmm\ model corresponds to a
$\Delta C=10.8$ statistic (rightmost column in Table~\ref{tab:poChandra}), 
compared to the best--fit model based on the
\chandra\ data. Given that the parent distribution of the $\Delta C$
statistic under the null hypothesis that the best--fit \xmm\ model is correct
is a $\chi^2(2)$
distribution, the null hypothesis can be discarded at a
$\geq 99.9$\% confidence level (i.e., the 99.9\% critical value is 9.2).
This statistic, however, does not mean that there is \emph{no} absorption
line at that redshift -- it simply means that the \chandra\ data suggests
a \emph{lower} parent value than the best--fit \xmm\ measurement. In practice,
this can be interpreted with the fact that the low signal--to--noise of the
\chandra\ data cannot conclusively confirm the \xmm\ detection, neither
completely rule out the absence of the line.

\subsection{Sources of systematic uncertainty}
\label{sec:systematics}

The results of the broad--band fits provided in Table~\ref{tab:results} show a \cmin\ statistics that consistently exceeds
the number of degrees of freedom, i.e., its expectation under the null hypothesis. 
The typical number of counts per 20~m\AA\ bin is always larger than $\sim$300 counts across
the entire 13-33\AA\ wavelength range, and therefore
the \cmin\ statistic is expected to be approximately distributed like a $\chi^2$ distribution. 
This asymptotic distribution is discussed in detail in \cite{kaastra2017} and in \cite{bonamente2020,bonamente2022book}, including
in the low--count case. In the case of the redshift system 9, corresponding
to the first of the two \cite{nicastro2018} lines, a $\chi^2(1477)$ distribution has an expected value and a standard deviation
corresponding to a range $1477\pm54$, with a one--sided 99.9\% critical value of 1651. Given that the measured best--fit statistic
exceeds the critical value, even the rather flexible \texttt{spline} continuum model appears insufficient to
provide a statistically acceptable fit. The flexibility of the
\texttt{spline} model is provided by the large number of
parameters used (twenty--one, in 1~\AA\ intervals between 13 and 33~\AA), two of which are reported in 
in Table~\ref{tab:spline} to illustrate typical uncertainties. Moreover, the two \texttt{hot} components that model the Galactic foregrounds provide a typical reduction in the best--fit by $\Delta C \simeq 50$ for three additional free parameters.

This mismatch between the data and the model can be interpreted either as an indication
that the model should be rejected, or that there are other sources of systematic uncertainty that have not been included in the analysis.
Given that the data are known to feature possible sources of systematic error, as indicated in Sect.~\ref{sec:x}, and that the
best--fit model generally follows the data well without broad--band systematic deviations, it is reasonable to attribute the
slightly higher--than--expected fit statistic with the presence of systematic sources of error that go above the Poisson counting
errors.

In the case of regression with Gaussian data, it is possible to account for systematic errors in a number of ways,
which typically result in an increase of the uncertainties above the Poisson or $\sqrt{n}$ errors.~\footnote{A review
of such methods can be found in Chapter~17 of \cite{bonamente2022book}.}
For a regression with the Poisson--based $C$ statistic,
those avenues are not available, given that the Poisson distribution enforces a variance that is always equal to its mean ---
in other words, the Poisson distribution does not have the flexibility to change its variance independently of its mean.
An alternative method to address sources of systematic error which is applicable to Poisson data consists of
considering an intrinsic model variance, whereby the additional source of uncertainty is attributed to the \emph{model}, and not the data.
A method to estimate the intrinsic model variability for Poisson data is presented in a companion paper,
where the statistical details of the method are described in detail \citep{bonamente2023}. For the present paper, it is sufficient
to report that these \xmm\ data are consistent with a systematic error of order few percent, meaning that the 
model is subject to variation of this order in each independent bin. This is consistent with the estimate of 
a few percent systematic error that was briefly discussed in Sect.~\ref{sec:x}, based on the knowledge of the instruments. Further details on the method  to estimate systematic errors can be found in \cite{bonamente2023}.

Fortunately, the narrow--band regressions provided in Table~\ref{tab:po}
 feature fit statistics that are in better agreement 
with the parent distribution. For example, the value of $C_{\text{min}}=119.7$ for 87 degrees of freedom, corresponding 
to the fit for the first \cite{nicastro2018} line at $z=0.4339$, has a 99.9\% critical value of 133.5 
according to the corresponding $\chi^2$ distribution.  This makes the measured
value of the fit statistic statistically consistent, at that level of confidence, with the model. Similar considerations apply for 
all the other fits presented in Table~\ref{tab:po}. Given that the conclusions presented in this paper are
based primarily on the results presented in that table, we do not expect that any possible sources of additional
systematic errors would have a significant impact on the results of this paper.

\def\ion{\mathrm{ion}}
\section{Cosmological implications of the search for X--ray absorption lines}
\label{sec:cosmology}

Of the \nFUV\ absorption lines systems detected by \cite{danforth2016} and used 
as sign--posts for possible X--ray absorption, there is only marginal evidence for
one new
detection
of \ovii\ (at redshift systems 1 and 2). Given that
the search was conducted using an FUV prior for the redshift, the statistical
significance of detection is the formal significance as obtained from the
detection statistics, without the need to account for redshift trials
(which are needed for blind searches, see e.g. \citealt{nicastro2005,nicastro2013,bonamente2019}).
For the other redshift systems, we only obtained upper limits to the non detection
of the \ovii, \oviii\ and \neix\ ions.

 This section
presents the methods to measure the cosmological density of the X--ray absorbing WHIM,
and it includes a new statistical framework to account for the limited sensitivity 
of the X--ray data, the use of FUV priors, and the presence of upper limits.
This new method is then applied to 
the possible detection of \ovii\ at $z=0.1876$ 
and alternatively
to the upper limits to the non--detections for the seven FUV--prior redshifts.
The result of this analysis is an estimate of the
cosmological density of WHIM associated with the detection or the non--detection of these
absorption line systems. It is first necessary to describe the statistical
methods used to derive the cosmological constraints, given the unique challenges posed
by the low--resolution X--ray data and the selection of the 
redshifts used to search for X--ray absorption lines.

\subsection{Statistical methods to constrain $\OWHIMX$}
The general method to constrain the cosmological density of baryons traced by
a specific ion was developed by \cite{tilton2012} and \cite{danforth2016}, whereby 
the distribution of the number of absorbers per unit column density and redshift,
\begin{equation} 
f(n)=\partial^2 n /\partial N \partial z,
\label{eq:fn}
\end{equation}
is used to evaluate the cosmological density
\begin{equation} \Omega_{\ion} = 
\dfrac{\rho_{\ion}}{\rho_c} = 
\dfrac{m_{\ion} H_0}{\rho_c c} \int f(n) N dN
\label{eq:Omegaion}
\end{equation}
where $N$ is the ion column density  and $\rho_c$ the critical density of baryons.
Given that the bulk of the baryons are expected to be in the neutral and ionized hydrogen, 
the ion is then
used as a tracer of the total baryonic matter via
\begin{equation}
 \Omega_{\mathrm{WHIM,ion}} = \Omega_{\ion} \dfrac{\mu_H \, m_H}{m_{\ion}} \dfrac{1}{A \cdot f_{\ion}(T)} 
\label{eq:Omega}
\end{equation}
with knowledge of the chemical abundance $A$ and temperature of the WHIM plasma. 
The mean atomic mass $\mu_H$ represents the mass associated with 
one hydrogen atom, which is a function of the state of ionization and chemical abundance of the absorber.
For a significantly sub--Solar chemical composition, we approximate it with $\mu_H=1.3$, similar to what
is assumed for other WHIM studies \citep[e.g..][]{nicastro2018}.
Equation~\ref{eq:Omega} is complicated by the fact that usually neither the chemical abundance
or the temperature are known accurately. To overcome these limitations, \cite{tilton2012}
assumed a temperature that corresponds to the peak of the ionization curve for the ion
under consideration, and a fixed value of $A=0.1$ Solar, while \cite{danforth2016} adopted 
an abundance that was motivated by simulations.

X--ray observations of the WHIM in absorption often yield only a few positive detections, and more 
often just upper limits
to the non-detection of a specific ion. Therefore, it is necessary to investigate an
alternative way of constraining the cosmological density of baryons, given that the
distribution \eqref{eq:fn} cannot be constrained effectively with only a few systems available.
A useful starting point is the simplification of Eq.~\ref{eq:Omega}
\begin{equation}
\OWHIMX= \dfrac{\mu \, m_H}{\rho_c} \dfrac{{\sum_i} N_{H,i}}{\sum_i D_i}
\label{eq:OmegaWHIMX}
    \end{equation}
    \citep[see, e.g.,][]{schaye2001,nicastro2018}
where the sum extends over all available sources, and the $i$--th source probes
a cosmological distance $D_i$ with a detection of a total WHIM column density $N_{H,i}$, representing
the entire baryonic content of the WHIM. 
It remains to be established how many X-ray absorbers, 
or how long a path length one needs to probe, so that one can derive reliable estimates for the cosmic baryon budget
that are applicable to the entire Universe. Obviously, measurements from one absorber alone, as in the
case of these data and those of \cite{nicastro2018}, is not adequate to make cosmological inferences. It is nonetheless
useful to develop this method, also for the sake of its applicability to future studies with a larger sample of sources.

When the estimate is provided by a single ion, e.g., \ovii\ or \oviii\ as is often the case
with X--ray observations,  
equation \eqref{eq:OmegaWHIMX} is equivalent to
\begin{equation} 
\begin{cases}
\Omega_{\mathrm{WHIM, ion}} = \Omega_{\text{ion}}  \dfrac{\mu \, m_H}{m_{\ion}} \dfrac{1}{A \cdot f_{\ion}(T)}\text{, with}
\\[10pt]
\Omega_{\text{ion}}= \dfrac{\rho_{\text{ion}}}{\rho_c} \text{, and}\\[10pt]
\rho_{\ion} = m_{\ion} \dfrac{N_{\ion}}{D_i},
\end{cases}
\label{eq:OmegaWHIMX2}
\end{equation}
where $N_{\text{ion}}$ is the column density of the specific ion under consideration and $\rho_c$ is the
critical density of baryons at the present epoch.
In these equation, the temperature and abundance is needed to convert
the ion's column density to the total plasma column density.  In principle, temperature constraints
can be obtained in the presence of at least two lines from different ions of the same element,
such as \ovi\ and \ovii\ (using, e.g., the ionization equilibrium curves of Fig.~\ref{fig:ions}),
and the abundances can be also constrained with additional lines from other elements. 
Such constraints are typically beyond the quality of the current X--ray data, and therefore it is customary 
to parameterize the resulting cosmological densities in terms of the unknown temperature (or
ion fractions) and abundances.

\subsection{Using \eagle\ to correct for sample selection and X--ray sensitivity limits}
The equations developed in the previous section need to be further
developed to account for (a) the limited sensitivity of X--ray observations that render a fraction
of the WHIM unobservable, (b) the selection of sightlines based on their
FUV priors, and also (c)  the presence of
upper limits to the non--detection. This section provides quantitative methods to 
address these issues.

\subsubsection{Definition of probabilities}
The sensitivity of the X--ray observations can be defined as an upper limit, e.g.  \Nosevenplus\ ,
with the meaning that the data at hand can only detect a column density of \ovii\ at that redshift
 than this larger than value, at a given level of confidence. Similar considerations
apply also to the other ions.
In other words, the available data are not sensitive to column densities \Noseven$\leq$\Nosevenplus,
due to the limitations of the instrument and exposure times.
In order to estimate the fraction of an ion's column density that in unobservable because of the limited
quality of the data, we make use of the  \texttt{EAGLE} simulations \citep{schaye2015,eagle2017}, leveraging the
prior relevant analysis by \cite{wijers2019}.
Figure~\ref{fig:joint}, reproduced from \cite{wijers2019}, illustrates the 
\texttt{EAGLE} prediction for the distribution of the column densities of \ovi\ and \ovii\ ions in a $100^3$~Mpc$^{3}$ simulation
of the Universe at low redshift, and for the distribution of \hi\ and \ovii. These joint distributions
are the basis for the prediction of the amount of \ovii\ present along the sightline. The same  analysis could be repeated for \oviii,
or other X--ray ions. For these ions, there is no information on line widths that was used for their selection,
and  therefore they represent the entire ionic budget in the \eagle\ simulations. 
The distributions of temperature and abundances  of the gas used for Figure~\ref{fig:joint} are described in detail in  \cite{wijers2019}. At $\log N_{\text{OVII}} (\text{cm}^{-2})>15$, the temperature of the gas is primarily in the range $\log T(\text{K})=5.5-6.5$, while at $\log N_{\text{OVII}} (\text{cm}^{-2})>16$
the temperatures are typically  $\log T(\text{K})=6.0-6.5$. The metallicity of the 
absorbers is almost entirely in the range $\log A = -1$ to $-0.5$ Solar.

We start by defining the probability $p_j$ that
a sightline with a detected \ovi\ column density $N_{\text{OVI}}$ at a specific redshift $z_j$ 
has an amount of \ovii\ larger than the upper limit,
\begin{equation} p_j = P(N_{\mathrm{OVII}} > N_{\mathrm{OVII}}^+ | N_\mathrm{OVI}).
\label{eq:pi}
\end{equation}
Such probability distribution is illustrated in Figure~\ref{fig:conditional}, which is obtained
from a vertical slice of  Fig.~\ref{fig:joint} at a given value of the \ovi\ column density.
Same considerations apply to \hi, whereby the probability in \eqref{eq:pi} is obtained by conditioning
on the column density $N_{\text{HI}}$ instead of $N_\mathrm{OVI}$.
Equation~\ref{eq:pi} reads as
the probability that the column density exceeds the upper limit, \emph{given} the 
value of the detected \ovi\ column density. 
Given that the systems under consideration are those with a known value of the \ovi\ and \hi\ 
column density,
the conditional probabilities of \eqref{eq:pi} illustrated in Fig.~\ref{fig:conditional} 
can be immediately used to evaluate the probability $p_j$ by evaluating the integrated probability
above the upper limit.
This probability quantifies what $p_j<1$ fraction of \ovii\ absorbers, by number, is observable
because of the limits in sensitivity
of the X--ray data. Usually this value is small,
indicating that the data 
can only detect a small fraction of possible \texttt{EAGLE}--predicted
\ovii\ absorbers. 

Since the only detectable absorbers (in the CGM and in filaments alike) are the ones with
the highest column density, it is necessary to evaluate another
probability that describes the \textit{fraction of the \ovii\ column density} detectable 
above the upper limit. This probability is defined as 
\begin{equation}
    P_j = \dfrac{\int_{N_{\mathrm{OVII}}^+}^{\infty} P(N_{\mathrm{OVII}}/N_\mathrm{OVI}) dN_{\mathrm{OVII}}}
        {\int_{0}^{\infty} P(N_{\mathrm{OVII}}/N_\mathrm{OVI}) dN_{\mathrm{OVII}}}
        \label{eq:Pi}
\end{equation}
where $P(N_{\mathrm{OVII}}/N_\mathrm{OVI})$ is the same conditional
probability distribution used in \eqref{eq:pi}, and it is weighed by the ion column density $dN_{\mathrm{OVII}}$.
This probability describes the cumulative fraction of \ovii\ column density 
that is detectable, and it is a larger number than $p_j$ because 
the larger column densities are the observable ones.
The evaluation of $P_j$ is illustrated in Fig.~\ref{fig:bias}, where the solid black
curve is the cumulative distribution of the marginal probability distribution in
Fig.~\ref{fig:conditional}, and the blue curve represents 
the fraction of \ovii\ column density present \textit{below} that value of $\log N_\mathrm{OVII}$, i.e.,
its cumulative distribution function.
For example, for $\log N_\mathrm{OVII}^+=15.3$, the curve in the left panel shows $p_j\simeq 0.36$, corresponding to
a fraction $P_j \simeq 0.72$  of \ovii\ \textit{above} that limit.
For the same source, these probabilities $P_j$ will in general vary with the
redshift of the \ovi\ priors, although variations are not expected to
be large if the X--ray data have a similar S/N throughout their redshift
coverage, as is the case for our data (see Table~\ref{tab:results}). 

In summary, the correction provided by $P_j$ according to \eqref{eq:Pi} probability follows the assumption that \ovii\ absorbers
are associated with \ovi\  (or \hi), and therefore we use the measured \ovi\ (or \hi) column density
to estimate the fraction of column density that is missed because of the limited
X--ray resolution. A similar analysis can be performed using different simulations, in order to address the dependence of the results  on the assumed distribution of the absorbing gas. While an in--depth comparison of these distributions among various simulations goes beyond the scope of this paper, we point out that the \texttt{CAMELS} cosmological simulations \citep{butler2023} feature a similar distribution of \ovii\ vs. \hi\ column densities (see their Figure 4) to the one from \texttt{EAGLE} used in this paper. In general, however, the distribution of column densities of ions are quite sensitive to the choice of simulation methods.

\subsubsection{Use of probabilities for cosmological estimates}
These probabilities can be now used as follows for obtaining cosmological constraints from the measured column densities.

(1) For a source (labeled by the index $i$) with one or more detections
(labeled by $j$), the cosmological
density can be constrained to
\begin{equation}
\rho_{\mathrm{ion,i}}= m_{\ion} \dfrac{\sum_j N_{\mathrm{ion,j}} /P_j}{D_i} 
    \simeq m_{\ion} \dfrac{\sum_j N_{\mathrm{ion,j}}}{D_i \cdot \overline{P_j}}
\label{eq:rhoion}
\end{equation}
where the sum extends to the number of redshifts with detections (for \es, this is only one,
the putative detection of \ovii\ at redshift system 1) and, 
in the limit of small differences in the
values of the $P_j$'s for that ion, the approximation holds with better accuracy when the average 
value $\overline{P_j}$ is used. 
The effect of $P_j$ is to reduce the  redshift path or, equivalently, the distance $D_i$ probed, 
to account for the fact that a fraction $(1-P_j)$ of the
total predicted \ovii\ column density was unobservable to begin with. The effect is therefore
to boost the cosmological density associated with
the actual detection, to recognize the inherent
difficulty to achieve the detection that was posed by
the limited sensitivity of the data.

(2) In the case of systematic non--detections of
an ion (i.e., \ovii) at all redshifts, 
the procedure must account for \emph{all upper limits} set along each sightline. 
In fact, the search for the WHIM was conducted at a number of redshifts 
$z_j$ with  $j=1,\dots,N_i$. In the case of \es,  $N_i=7$ 
independent redshifts were being tried. 
For each of these redshift trials, EAGLE
predicts a distribution of X--ray column densities according to a curve of the type of
Figure~\ref{fig:conditional}, based on the positive detection
of an FUV ion (either \ovi\ or \hi). Such systematic non--detection of
X--ray absorption can be taken to mean
that none of the \emph{observable} WHIM was present at any redshift where
EAGLE predicted it. This situation can be quantified in the following manner.

First we calculate the \emph{expectation} of the \emph{undetectable} $\nionj$ as
\begin{equation}
    \E[\nionj^+] = \dfrac{\int_{0}^{\nionj^+} N_\mathrm{ion} P(N_\mathrm{ion}) dN_\mathrm{ion}}
    {\int_{0}^{\nionj^+} P(N_\mathrm{ion}) dN_\mathrm{ion}},
    \label{eq:ENionPlus}
\end{equation}
which is a probabilistic expectation or average column density of what XMM
could have missed at that redshift, given the presence of an FUV ion. In this equation,
$P(N_\mathrm{ion})$ is the same distribution used in \eqref{eq:pi}, but the conditioning on
the FUV ion was omitted for the sake of clarity. In other words, this is the average value of 
the column density of an X--ray ion as predicted by EAGLE for that redshift trial, 
but that are not detectable due to the sensititivity of the
X--ray data; the denominator ensures that the probability distribution
is properly normalized.

Such average values can be summed over all non--detections, so that
\begin{equation}
    \E[\nion^+]=\sum_{j=1}^{N_i} \E[\nionj^+]
    \label{eq:nionMinus}
\end{equation}
represents an \emph{average upper limit} to the column density of
the ion that was systematically not detected. The use of \eqref{eq:nionMinus}
to estimate the overall upper limit to a systematic non--detection
has two main advantages. First, it accounts for all redshift trials; and
second, it makes use of both the X--ray sensitivity limits and 
the expected distribution of X--ray ions according to EAGLE.

This average upper limit can then be used in
\begin{equation}
\rho_{\ion}  \leq m_O
\dfrac{\E[\nion^+]}{D_i}
\label{eq:rhoionUL}
\end{equation}
to provide an upper limit to the volume density of the ion, in a manner
similar to \eqref{eq:rhoion} in the case of a detection. Accordingly, 
an upper limit to the associated cosmological baryon density of the WHIM is obtained 
via \eqref{eq:OmegaWHIMX2}, which requires assumptions
on the temperature/ionization fraction, and overall abundance of the
element under consideration (i.e., typically oxygen). 

It is worth pointing out that, when a sightline such as \es\ has several undetected
X--ray lines, $\E[\nion^+]$ is the sum of several terms that increase the
average upper limit, above that of a single upper limit. 
This is reasonable, in that each independent redshift trial represents an independent opportunity,
according to the \texttt{EAGLE} predictions, for the presence of associated X--ray ions.
An alternative to \eqref{eq:nionMinus} would be to simply use the least--stringent X--ray upper limit $\nionj^+$ in place of $\E[\nion^+]$, but such procedure would
not account for all redshift trials included in the distance $D_i$ probed by the
X--ray data.


\subsection{Cosmological constraints from \es}
\label{sec:cosmoConstraints}
We now illustrate the method provided in the previous section with
the column densities measured from the \es\ \xmm\ data.
We entertain two scenarios. First, assuming that the \ovii\ line associated with the $z=0.1876$
absorber is a real astrophysical signal, despite its limited significance of detection,
we use \eqref{eq:OmegaWHIMX2} to estimate the cosmological significance
of the putative detection. Then, we use a more conservative approach to estimate an upper limit
to the systematic non--detection of X--ray ions, according to \eqref{eq:rhoionUL}. 
 We use a value for the sensitivity  of the \xmm\ spectra to the detection of the \ovii\ ion
as $\log N_\mathrm{OVII}^+ \simeq 15.3$, as discussed in  Sect.~\ref{sec:stack}.

For the $z=0.1876$ absorber, we consider an associated O~VI column density that corresponds to the sum of the
first two FUV lines in Table~\ref{tab:danforth}. In fact, these two FUV lines are indistinguishable at the resolution of these \xmm\ data, and 
the resolution of the \eagle\ distributions (see Fig.~\ref{fig:joint}) is such that those two absorbers would have been counted as one \citep{wijers2019}.  
The sensitivity of the X--ray data at hand corresponds to the ability of observing
a fraction $P_j \simeq 0.72$ of all the available \ovii\ along the sightline, give the presence of $\log N(\text{O~VI}) \geq 14.1$ at that redshift, according to \eqref{eq:Pi}. 

\subsubsection{The effective redshift path probed}

The uncertainties regarding the redshift of \es\ discussed in Sect.~\ref{sec:redshift} must be reflected in
the cosmological constraints from the FUV--prior search for X--ray absorption lines that we have conducted. 
The most likely redshift for the source is $z \simeq 0.433$, with an upper limits of approximately $z \leq 0.48$ \citep{nicastro2018}. Moreover, our choice of the redshift systems to investigate (see Table~\ref{tab:danforth})
is limited by the selection effects of the \cite{danforth2016} study, including the completeness of the 
\hi\ and \ovi\ systems (see Table~4 in \citealt{danforth2016}). Given these uncertainties, we conservatively assume a maximum redshift path of $z\leq 0.48$
for our search of X--ray absorption lines associated with the WHIM. If the $z \simeq 0.433$ is correct, then our estimates
of the cosmological density of baryons associated with the lines will be strict lower limits.

Moreover, the redshift path probed by the \xmm\ data is smaller
than the entire $z \leq 0.48$ path up to the maximum estimated redshift of the source.
Specifically, several Galactic oxygen lines (at $21.6-23.5$\AA, see Table~\ref{tab:Lines}) prevent an accurate blind search
for \ovii\ at $z \leq  0.09$, and several regions with poor calibration result in an estimated $\sim 25$\% reduction in 
the redshift path probed, as discussed in Sec.~\ref{sec:redshiftTrials}. 
The effective distance probed $D_i$ in \eqref{eq:OmegaWHIMX2} must therefore be evaluated accordingly.
For a flat $\Lambda$CDM cosmology with a dimensionless Hubble constant of $h=0.7$ and a critical
matter density at the present epoch of $\Omega_m=0.3$, the angular diameter distance
to $z=0.48$ is $D_A=1.23$~Gpc. The purpose of the ratio $N_{\text{ion}}/D_i$ in \eqref{eq:OmegaWHIMX2}, however,
is that of estimating a baryonic density associated with the measured
column density. The measured column density $N_{\text{ion}}$ is independent of the cosmology used,
since it is the result of the \texttt{slab} model applied to the spectrum, with no explicit dependence on
the parameters of the cosmological model. It therefore appears appropriate to estimate the effective distance 
$D_i$ via the comoving distance, which is a function of the comoving volume $V_C(z)$ associated with the redshift path available for the search,
instead of the angular diameter distance.
In this case, this effective distance can be estimated as
\begin{equation}
D_{\text{eff}} \simeq \left( A \times \dfrac{3}{4 \pi} \left(V_C(z_{\text{max}})-V_C(z_{\text{min}})\right) \right)^{\nicefrac{1}{3}}
\label{eq:Deff}
\end{equation}
with $z_{\text{max}}=0.48$ and $z_{\text{min}}=0.09$, and a correction factor of $A=0.75$ to account for the loss of
volume due to gaps in the data, for a value of $D_{\text{eff}} \simeq 1.65$~Gpc. 

It should not be surprising that this
effective distance based on the comoving volume is different, and significantly in excess of, the angular diameter distance. 
In fact, 
in a Friedmann--Lemaitre expanding universe with the usual Robertson--Walker line element, 
the angular diameter distance simply represents the ratio of an object's physical size to its angular size,
while the comoving volume is a volume measure in which the number density of non--evolving systems, such as the WHIM, remains of
constant density with redshift. It is clear that, when applied to cosmologically significant redshifts
as in the present case, Equation~\ref{eq:OmegaWHIMX2} needs to be modified to reflect the expansion and dynamics of
the universe. Eq.~\ref{eq:Deff} provides the means to obtain a simple approximation that accounts 
for the dynamics of an expanding Universe, and therefore
in the following we use \eqref{eq:OmegaWHIMX2} by replacing $D_i$ with 
the comoving volume--estimated $D_{\text{eff}}$. 

An alternative means to account for the dynamics of an expanding universe is to use the so--called
\emph{absorption distance} $X(z)$, defined by
\[ 
X(z) = \int_0^z \dfrac{(1+z)^2}{E(z)} dz
\]
where $E(z)$ is the usual evolution function. 
The absorption distance is a dimensionless quantity that represents an equivalent redshift path, 
and was introduced by \cite{bahcall1969} to study the probability of intercepting objects with non--evolving density
in the spectra of QSOs. Accordingly, one can define an equivalent distance for the search of absorption lines via
\begin{equation}
    D_X=A\times \dfrac{c\; \Delta X}{H_0}
    \label{eq:DX}
\end{equation}
where the same efficiency factor $A$ is assessed as in \eqref{eq:Deff}. For the same values of the minimum and maximum 
redshifts and of the cosmological parameters, \eqref{eq:DX} yields $D_X=1.79$~Gpc, i.e., 8\% larger than $D_{\text{eff}}$
from \eqref{eq:Deff}. This method was used, for example, by \cite{ryan2009} for the study of high--redshift absorption lines.
The use of $D_X$ according to \eqref{eq:DX} in place of $D_{\text{eff}}$ in \eqref{eq:OmegaWHIMX2} would have the effect of \emph{reducing} the estimates
of the cosmological density of baryons associated with the WHIM by less than 10\%.

A third estimate of $\Omega_{\mathrm{ion}}$ can be obtained using a directly inferred comoving ion density. A column density is simply the number of ions per unit physical surface area in an absorber. We can therefore infer the comoving ion density contributed to the total by absorber $i$ by first dividing the column density by $(1 + z_i)^2$, where $z_i$ is the redshift of the absorber. This yields the number of ions per unit comoving surface area for the absorber. By then dividing this by the comoving distance probed by the sightline, we get 
\begin{equation}
\rho_{\mathrm{ion}, C} = m_{\mathrm{ion}} \frac{\sum_i N_{\mathrm{ion}} / (1 + z_i)^2}{A (D_C(z_{\max}) - D_C(z_{\min}))},
\label{eq:rhodirect}
\end{equation}
where $\rho_{\mathrm{ion}, C}$ is the average comoving ion density along the line of sight and $D_C$ is the comoving distance. This yields an equivalent distance of
\begin{equation}
D_{\mathrm{direct}, i} = A (1 + z_i)^2 [D_C(z_{\max}) - D_C(z_{\min})]
\label{eq:Ddirect}
\end{equation}
for each absorber $i$ along the sightline. For our possible detection of \ovii\ at $z=0.1876$, \eqref{eq:Ddirect}
yields an equivalent distance of $D_{\mathrm{direct}, i} =1.53$~Gpc.

We note that this distance is related to the distance $D_X$ from eq.~\ref{eq:DX}. We can compare
\begin{equation}
D_{\mathrm{direct}, i} = A (1 + z_i)^2 
\int_{z_{\min}}^{z_{\max}} dz \frac{c}{H(z)},
\end{equation}
where $c$ is the speed of light and $H$ is the Hubble factor, to 
\begin{equation}
D_X = A \int_{z_{\min}}^{z_{\max}} dz (1 + z)^2 \frac{c}{H(z)}.
\end{equation}
The difference is whether the factor of $(1+z)^2$ is averaged over the redshift path ($D_X$) or applied to the individual absorbers ($D_{\mathrm{direct}, i}$). While eq.~\ref{eq:Ddirect} gives average comoving ion density along the line of sight directly, with eq.~\ref{eq:DX}, it is inferred from the total column density under the assumption that the absorber properties (in physical units) do not depend on redshift, and that the absorber population therefore only evolves through cosmological expansion.

In the following, we use the comoving distance as a measure of the effective distance path probed, with
$D_{\mathrm{eff}}=1.65$ Gpc according to \eqref{eq:Deff}. 
The use of the absorption distance $D_X$ according to \eqref{eq:DX}, in place of $D_{\text{eff}}$, would imply
the assumption that the WHIM is non--evolving with redshift, and have the effect of \emph{reducing} the estimates
of the cosmological density of baryons associated with the WHIM in \eqref{eq:OmegaWHIMX2} by less than 10\%.
Uncertainties associated with the calculation of the effective distance path are therefore 
estimated at the level of $\pm$10\%,
based on the analysis provided in this section.

\subsubsection{Estimate of baryons density assuming a detection of \ovii\ for the $z=0.1876$ \ovi\ absorber} 
\label{sec:ULResults}
Assuming a detection of \ovii\ associated with the $z=0.1876$ \ovi\ absorber,
with a column density of $\log N(\mathrm{O~VII})=15.13\pm^{0.19}_{0.31}$~cm$^{-2}$, according to Table~\ref{tab:results},
equations 
\eqref{eq:OmegaWHIMX} and \eqref{eq:rhoion} provide an
estimate of the mass density of \ovii\ of 
\[ \Omega_{\text{OVII}} = \dfrac{\rho_{\mathrm{OVII}}}{\rho_c} = (22 \pm 11) \times 10^{-6}, \]
and a corresponding cosmological density of
\begin{equation} 
\OWHIMX = 0.021 \pm 0.011 \left(\dfrac{A}{\text{0.1 Solar}}\right)^{-1}
\left(\dfrac{f_{\text{ion}}}{1.0} \right)^{-1}
\label{eq:OmegaResult}
\end{equation}
where a 10~\%\ Solar abundance of oxygen and a temperature
corresponding to an \ovii\ ion fraction  of 100\% were assumed as the reference values,
although they are free parameters that are not constrained by the data.
When compared with the standard expectation of $\Omega_b \simeq 0.048$ for the entire
baryon budget, we obtain
\begin{equation}
    \dfrac{\OWHIMX}{\Omega_b} = 0.44\pm0.22.
\end{equation}
It is therefore clear that a positive
detection of just one X--ray ion towards \es\ may be sufficient, in principle,
to account for the missing baryons at low redshift. However, a larger number of 
detections and a longer redshift path are necessary to overcome the uncertainties
associated with small--number statistics, as further discussed in Sect.~\ref{sec:systematicsCosmo}.

In principle, a more accurate estimate of $\OWHIMX$\ can be provided
by using all available X--ray information to constrain the temperature
of the WHIM. This could be done, for example,
by means of the \texttt{hot} model in \spex, which provides an overall estimate of the 
total $N_H$ column density, thereby enabling the use
of \eqref{eq:OmegaWHIMX} and requiring only the assumption of
an average metal abundance of all elements contributing to the estimate
(primarily oxygen and neon), instead of assuming a temperature and the corresponding ion fraction.
For these data, the limited significance of detection makes such additional modelling unwarranted,
and we do not attempt it in this paper.
We therefore simply parameterize our answer in \eqref{eq:OmegaResult} in terms of an assumed ion fraction for \ovii,
corresponding approximately to the peak value near $\log T(K)=6-6.4$, according to Fig~\ref{fig:ions} in collisional
ionization equilibrium.
This temperature range includes the median temperature within the densest
\emph{Bisous} filaments in \eagle, where most of the missing baryons are expected to be located
\citep[see, e.g., Fig. 13 in][]{tuominen2021}.
If the putative X--ray/FUV absorber at $z=0.1876$ has a lower temperature, as 
could be the case based on its strong FUV absorption, then this estimate of $\OWHIMX $ would increase accordingly.

\subsubsection{Estimate of the baryon density assuming systematic non--detections}
On the other hand, given the limited significance of the \ovii\ line 
at $z=0.1876$, it would be prudent to consider the scenario that none of the 7 FUV redshift systems
has significant \ovii\ detected.
Such systematic non--detection of \ovii\ at all redshifts with
\ovi\ and \hi\ BLA FUV priors can then be used to set an
upper limit to $\Omega_{\text{WHIM,X}}$. 
According to \eqref{eq:rhoionUL}, with an estimated sensitivity of
$\log N_{OVII}^+=15.3$ for the 7 independent \ovii\ redshifted lines probed, 
the systematic non--detection of \ovii\ leads to 
\begin{equation} 
\OWHIMX  \leq 0.043 \left(\dfrac{A}{\text{0.1\, Solar}}\right)^{-1}
\left(\dfrac{f_{ion}}{1.0} \right)^{-1}.
\label{eq:ULResults}
\end{equation}
In this estimate, the expectation or average value of the column density that is 
missed according to \eqref{eq:ENionPlus} is in the range 
$\E[N_{\text{OVII}}^+] = 5-7.6 \times 10^{14}$~cm$^{-2}$ for the three redshifts
with \ovi\ priors, and in the range of $0.4-1\times 10^{14}$~cm$^{-2}$ for the 
four redshifts with \hi\ priors. The lower expected \ovii\ column density in correspondence
of \hi\ BLA detections is understood according to Fig.~\ref{fig:joint}, whereas the correlation
between \ovii\ and \hi\ is lower, and in general \hi\ predicts lower amounts of X--ray absorbing WHIM. Notice that the distribution of ion densities in Fig.~\ref{fig:joint} includes all ion velocities, and therefore we do not specifically correlate
\ovii\ with \hi\ BLA systems, but rather with \hi\ in general, according to the analysis of \cite{wijers2019}.
The corresponding cosmological density is
\begin{equation} 
\dfrac{\OWHIMX}{\Omega_b}  \leq 0.90 \left(\dfrac{A}{\text{0.1\, Solar}}\right)^{-1}
\left(\dfrac{f_{ion}}{1.0} \right)^{-1}.
\label{eq:ULResults2}
\end{equation}

It is necessary to emphasize that the results in \eqref{eq:ULResults} are heavily dependent on
the expected distributions of \ovii\ according to the FUV priors, as measured by the \texttt{EAGLE} simulations.
It goes beyond the scope of this paper to investigate changes in these distributions
as a function of simulation parameters, and other aspects in the analysis of the simulations
leading to the joint distributions of Fig.~\ref{fig:joint}. The estimate presented in this section
should thus be considered an order--of--magnitude estimate, primarily aimed at illustrating the method of analysis
to obtain upper limits to the systematic non--detection of X--ray ions, while accounting for 
sensitivity of the instrument and redshift selection.

Assuming that \oviii\ and \neix\ are associated with the same WHIM
phase, it is sufficient to evaluate an upper limit to the
cosmological density of the WHIM using the ions with the most
stringent upper limits, in this case \ovii.
It is useful to point out that the upper limit to the systematic non--detection
of X--ray ions provided in \eqref{eq:ULResults} accounts for all
redshift trials with FUV priors, and uses the expectation for the
ions that were not detected, according to the \texttt{EAGLE} distribution of ions.

\subsection{Sources of systematic error in the cosmological analysis}
\label{sec:systematicsCosmo}
The method of analysis leading to the cosmological constraints
that was provided in this Section 
introduces several sources of uncertainties, in addition to
the uncertainties associated with the measurement of the column densities
that were discussed in Sect.~\ref{sec:systematics}. In the following we discuss some of the most
important sources of error.

\subsubsection{Temperature of the WHIM}
The key source of uncertainty in the estimates of the cosmological
density of baryons in \eqref{eq:OmegaResult} and \eqref{eq:ULResults} is
associated with the temperature of the absorber.
An estimate of the temperature, under the assumption of
a single--phase medium, can be obtained combining the \ovi\ and \ovii\ data.
Such approach was followed, e.g., by \cite{nicastro2018} for their
serendipitous \ovii\ line, whereby the assumption of collisional equilibrium
can be used to convert a column density ratio to a temperature.
The large uncertainties in the
only possible \ovii\ detection available are too large to warrant such analysis in this case.
In the present analysis, we simply parameterized the cosmological density as a function of the unknown
ionization fraction, which is a function of temperature according to the curves in Fig.~\ref{fig:ions},
under the assumption of collisional equilibrium. If the temperature is near the peak of the \ovii\ ionization,
$\sim \log T(K)=6-6.4$, then the uncertainties in the estimates \eqref{eq:OmegaResult} and \eqref{eq:ULResults} 
are of order a few times 100\%, since in this range of temperatures the ionization fraction can vary by a factor of few.
Should the temperature be significantly lower, as is suggested by the strong \ovi\ absorption, the estimates
would increase by an order of magnitude, or more. It is therefore clear that an estimate of the absorber's temeperature
is a key factor in drawing cosmological estimates from X--ray column density measurements.

\subsubsection{Chemical abundance of the WHIM}
Even in the presence of a temperature estimate, the abundance of the absorber also
remains unknown. It is customary to rely on observational estimates of metal abundances in the
outskirts of clusters, or on theoretical estimates for the WHIM, which consistently
indicate a significantly sub--Solar abundance for the inter--galactic medium \citep[e.g.][]{mernier2017, cen2006, pearce2021}.
Uncertainties associated with such estimates are typically of the order of few times 100\%, rendering the 
associated cosmological estimates uncertain by the same factor.

\subsubsection{Use of FUV priors}
Another source of systematic uncertainty already alluded to in Sect.~\ref{sec:ULResults} is the estimate of
probabilities of detection based on the \cite{wijers2019} analysis of the \texttt{EAGLE} simulations.
Our method consists in using the measurement of FUV column densities as a prior for the
distribution of \ovii\ column densities. An extreme example of the impact of this assumption in the
estimate of the $\OWHIMX$ parameter according to \eqref{eq:OmegaResult} is if we ignore this probability altogether,
effectively setting 
it to a value of $P_j=1$ in \eqref{eq:rhoion}. In this case, the $z=0.1786$ \ovii\ line would correspond to a measurement of 
\begin{equation*} 
\OWHIMX = 0.016 \pm 0.009 \left(\dfrac{A}{\text{0.1 Solar}}\right)^{-1}
\left(\dfrac{f_{\text{ion}}}{1.0} \right)^{-1}
\label{eq:OmegaResult}
\end{equation*}
which, as expected, is lower than the results in \eqref{eq:OmegaResult}. This estimate corresponds to the assumption that
the \xmm\ data have the ability to detect all X--ray absorbing WHIM along the sightline.

\subsubsection{The fair--sample assumption and small--number statistics}
Finally, the  assumption underlying the identification of the missing baryons via WHIM filaments is that
the sightlines under consideration for the estimate, in this case a single line towards a quasar at $z \simeq 0.48$, is
\emph{representative} of the cosmos at large. With seven independent redshift trials based on the FUV priors, 
each approximately with a probability of a detection of X--ray signal (above the sensitivity level) of $\sim 30$\%,
the estimate is clearly dominated by small--number statistics. For comparison, the \cite{tilton2012}
estimate of the cosmological density of the FUV--absorbing WHIM was based on 44 sightlines with an \hi\ Lyman--$\alpha$
redshift path of  $\Delta z=5.38$ and an \ovi\ redshift path of $\Delta z=6.1$,
and the \cite{danforth2016} estimate used 82 sightlines with an \hi\ Lyman--$\alpha$
redshift path of  $\Delta z \simeq 20$ and an \ovi\ redshift path of $\Delta z=14.5$. 
Such small--number statistics can  overestimate the cosmological estimate
of WHIM baryons, if, e.g., a sightline with a short redshift path has one or more WHIM systems detected,
as was the case for the original \cite{nicastro2018} results. This could be the 
result of a chance
alignment with a filament with high column density of absorbing material. This situation can only be ameliorated
with the investigation of longer redshift paths and with more sources at different positions in the sky.

In consideration of all these sources of systematic uncertainties, the
estimates provided in this paper for the \es\ sightline should be regarded as order--of--magnitude
estimates. More accurate estimates require, at a minimum,
a larger number of sources for a correspondingly larger redshift path, 
and an accurate knowledge of the temperature of the absorbers.
According to Eq. \ref{eq:rhoionUL}, the upper limit to the cosmological density associated to the WHIM is inversely proportional to the effective distance probed, for sources with observations of comparable depth. 
We therefore estimate that a redshift path--length of order \emph{ten times larger} than what is available with the present data is needed to provide upper limits that would challenge the hypothesis that the missing baryons are in a warm--hot phase of the inter--galactic medium. In addition to more data, other sources of systematic errors, especially with regards to temperature and abundances, would also have to improve dramatically with respect to the current grating spectrometry data. Such improvements would be provided by upcoming missions such as the proposed \emph{Line Emission Mapper} \citep[\emph{LEM},][]{LEM2022}, with sufficient spectral resolution to constrain the temperature and metal abundance of the absorbers, in addition to an increased sensitivity to the intervening column density.

\subsection{Comparison with prior 
results for the cosmological density of the WHIM}

Our results for the cosmological density of the X--ray absorbing WHIM compare well with a similar investigation
performed by \cite{kovacs2019} using \emph{Chandra} data of the source $H~1821+643$ at $z=0.297$.
In their analysis, the stacking of 17 redshift systems with FUV priors resulted in a $3.1$--$\sigma$
detection of a possible \ovii\ He--$\alpha$ line, which corresponds to a cosmological density of
\[
\OWHIMX=0.023\pm0.007,
\]
if an abundance of 10\% Solar is used.
Their measurement therefore corresponds to a value of 
\[ \dfrac{\OWHIMX}{\Omega_b}=0.48\pm0.15,
\]
which is consistent with our results based on the possible detection
of \ovii\ at $z=0.1876$ in \es. It is also useful to note that the \texttt{CAMELS} analysis of \cite{butler2023} indicates that the WHIM column densities implied by the \cite{kovacs2019} results are significantly larger than the prediction from numerical simulations. While it cannot be ruled out that a single sightline might have an unusually large column density of \ovii\ and other ions, this possible discrepancy further highlights the need for a larger sample of sources from which to make cosmological inferences.

Although there is significant evidence that the  \cite{nicastro2018} detection
does not originate from the WHIM (see Sect.~\ref{sec:redshift}), we nonetheless
compare our analysis of the putative \ovii\ $z=0.4339$ line with their cosmological implications.
%
If we apply the same cosmological equations as used by \cite{nicastro2018}, i.e., Eq.~\ref{eq:OmegaWHIMX2},
to our measurement of the \ovii\ column density for their serendipitous line at $z=0.4339$ ($\log N(\text{OVII})=15.74\pm0.11$),
with the same comoving distance as estimated for the FUV--prior lines,
we infer a cosmological density of 
\begin{equation}
    \OWHIMX = 0.031 \pm 0.008 \left(\dfrac{A}{\text{0.2 Solar}}\right)^{-1}
\left(\dfrac{f_{\text{ion}}}{1.0} \right)^{-1}
\end{equation}
corresponding to a fraction $65\pm17$\% of the \planck\ baryon density. Our estimate is therefore
equivalent to the one reported in \cite{nicastro2018}, and consistent with their conclusion that
this absorption line feature, if interpreted as \ovii\ He--$\alpha$ at $z=0.4339$, is
consistent with the hypothesis that the hotter portion of the X--ray absorbing WHIM constitutes the missing
baryons.

\begin{figure*}
\includegraphics[width=3.2in]{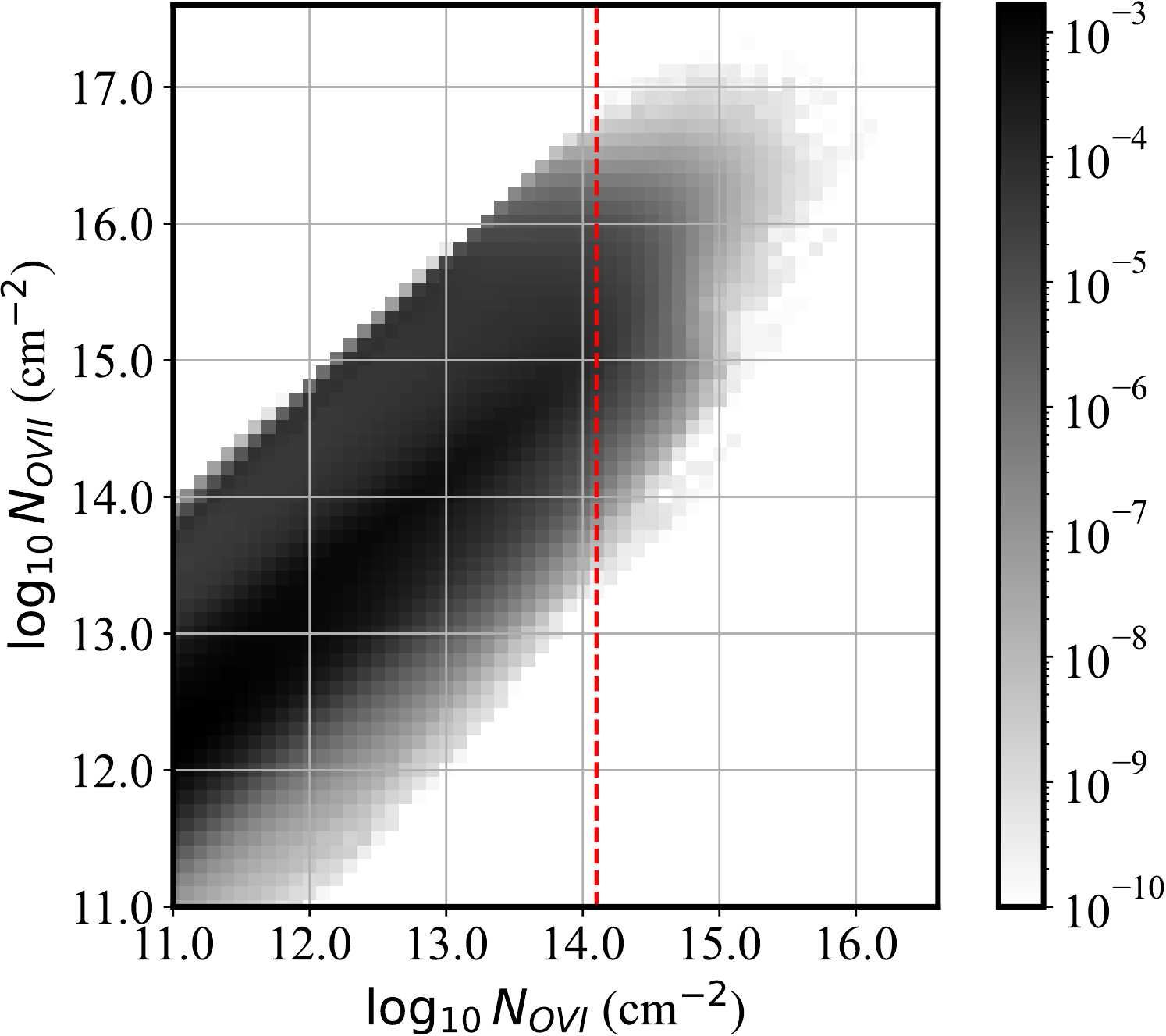}
\includegraphics[width=2.8in]{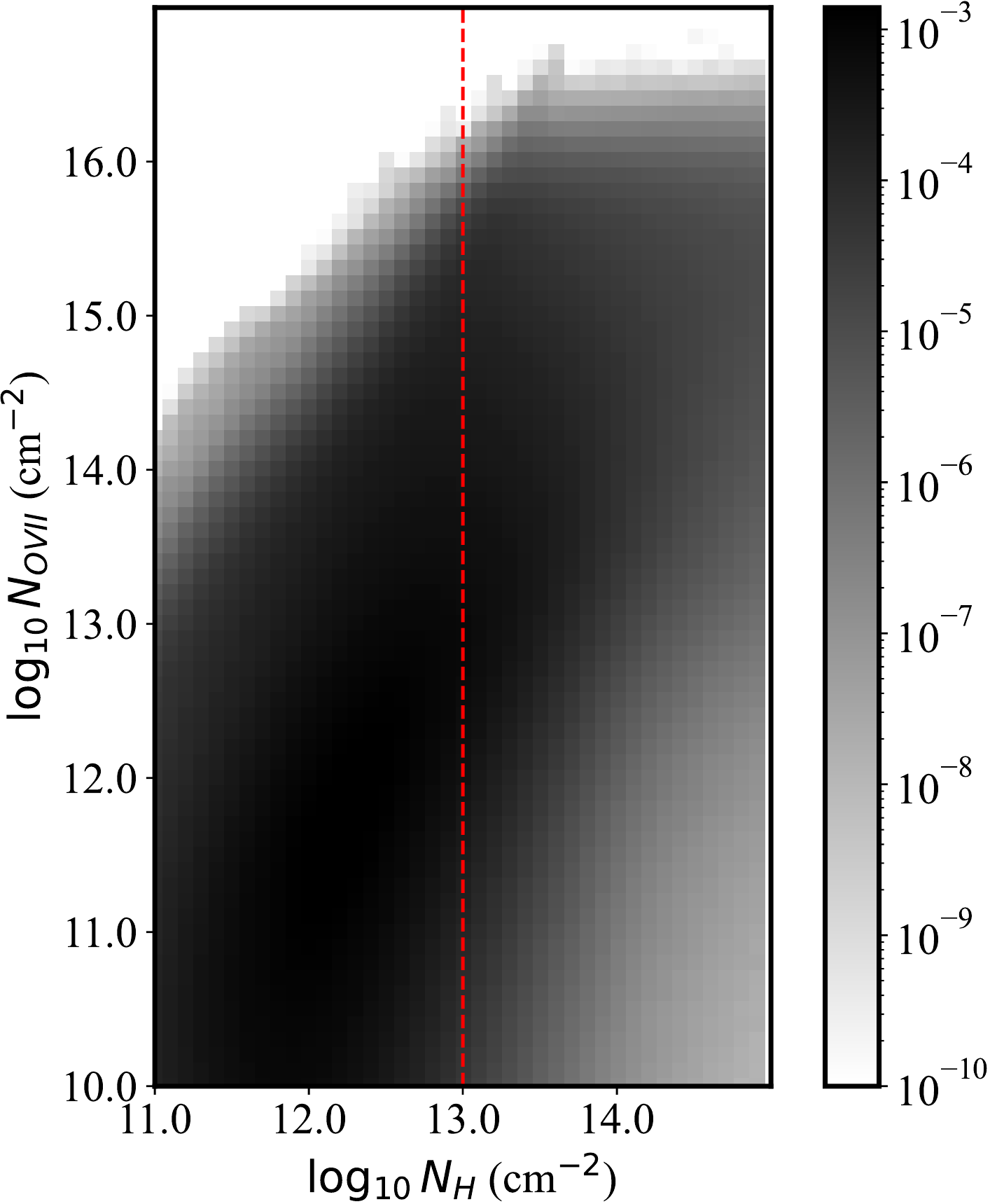}
\caption{Joint probability distribution of (a) \ovii\ and \ovi, and (b) \ovii\  and \hi,
from the \eagle\ simulations of \protect\cite{wijers2019}. The red lines
represent respectively the value of the combined \ovi\ column density detected for absorbers 1 and 2 ($\log N(\mathrm{O~VI})=14.1$),
which feature a possible \xmm\ detection of \ovii, and the value of a characteristic \hi\ BLA detection
for systems 5--8. The distribution
represents the number of sightlines in \eagle\ with those column density values,
and therefore they are not normalized.}
\label{fig:joint}
\end{figure*}

\begin{figure*}
\includegraphics[width=3.2in]{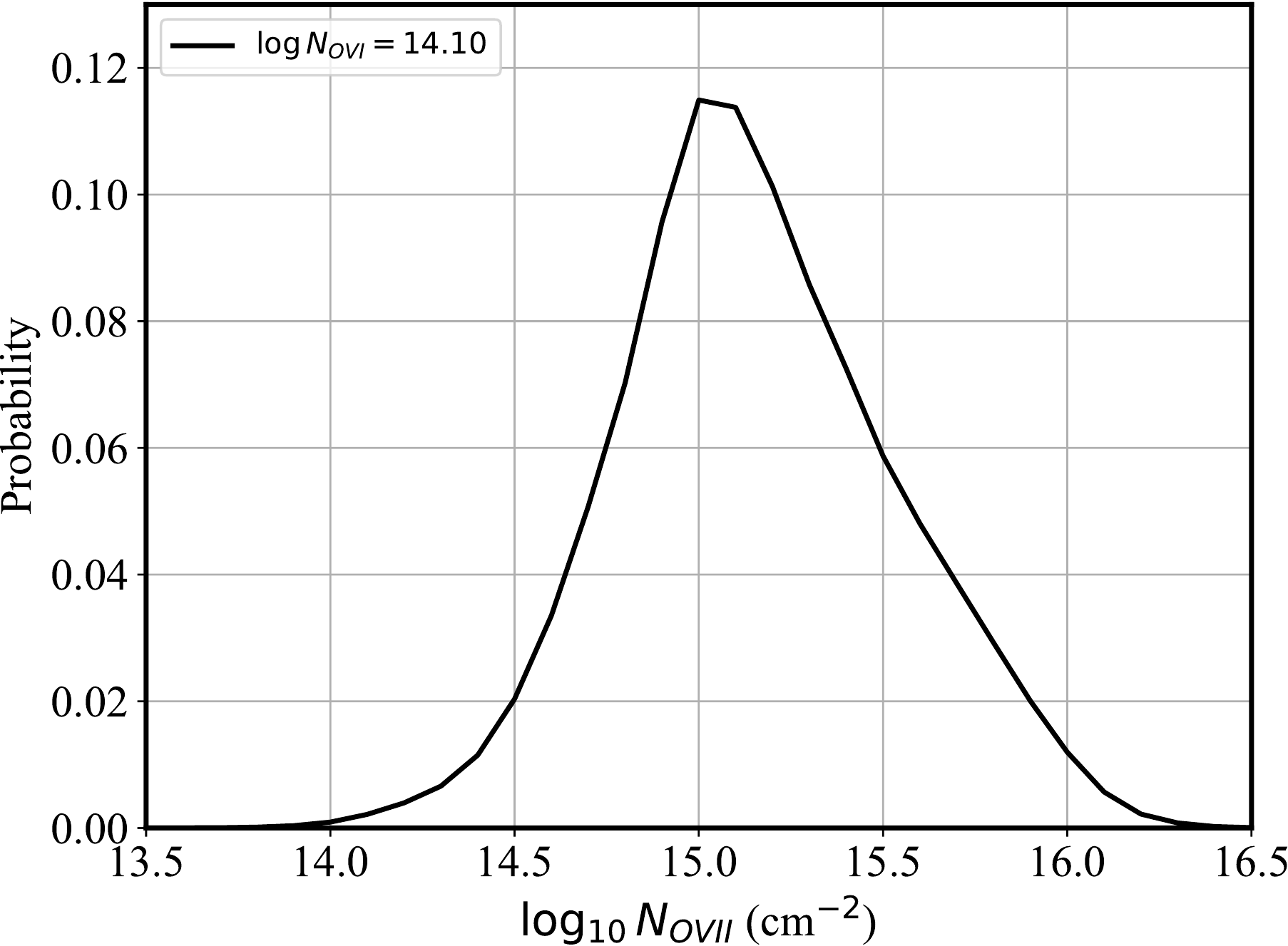}
\includegraphics[width=3.2in]{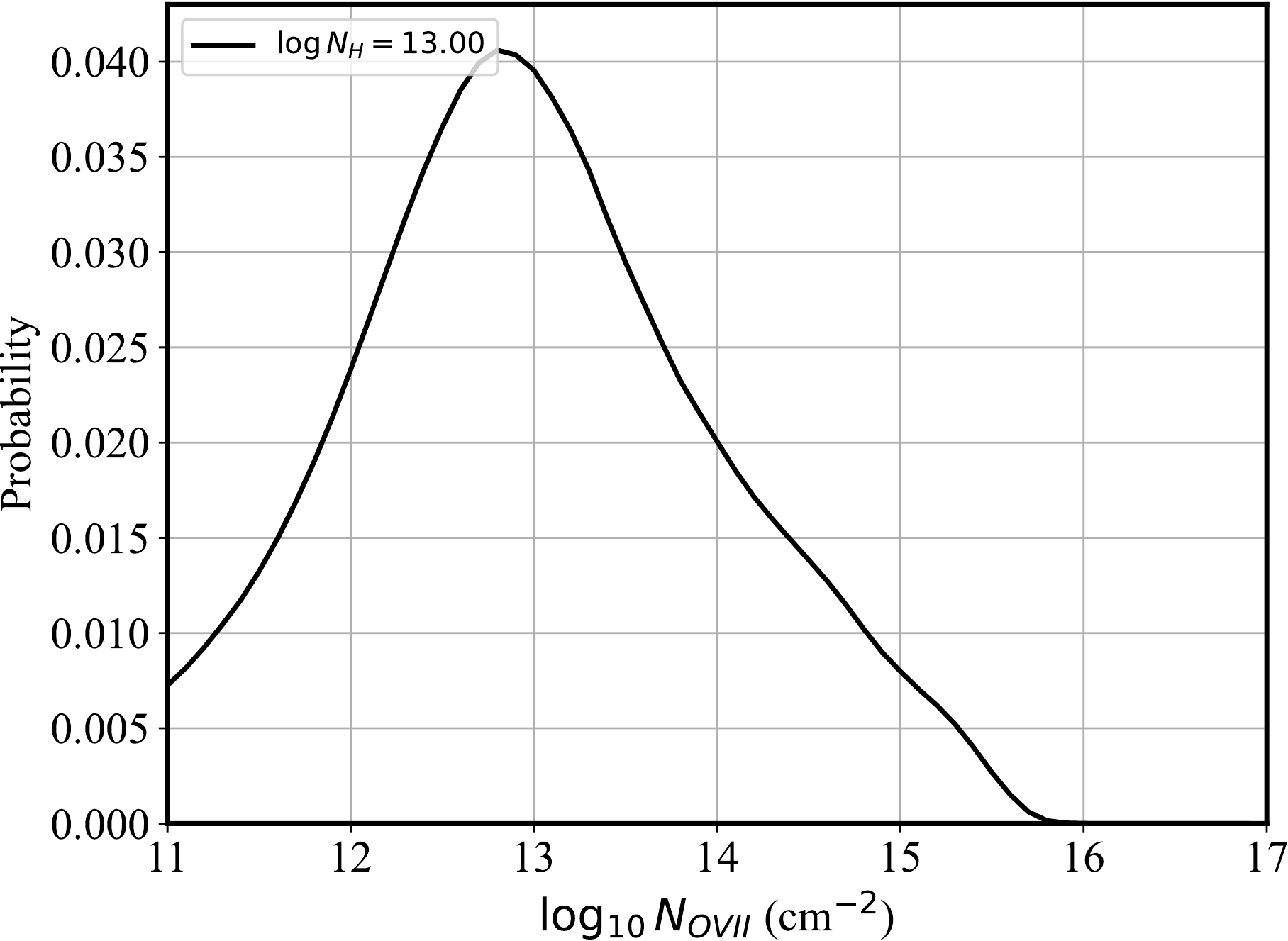}
\caption{Conditional probability distribution of (a) \ovii\ given a range of \ovi, 
  and (b) \ovii\  given \hi,
from \protect\cite{wijers2019}. The distributions correspond to the solid red lines
in Fig.~\ref{fig:joint}, and have been normalized to a total probability of one.}
\label{fig:conditional}
\end{figure*}

\begin{figure*}
    \centering
    \includegraphics[width=3.2in]{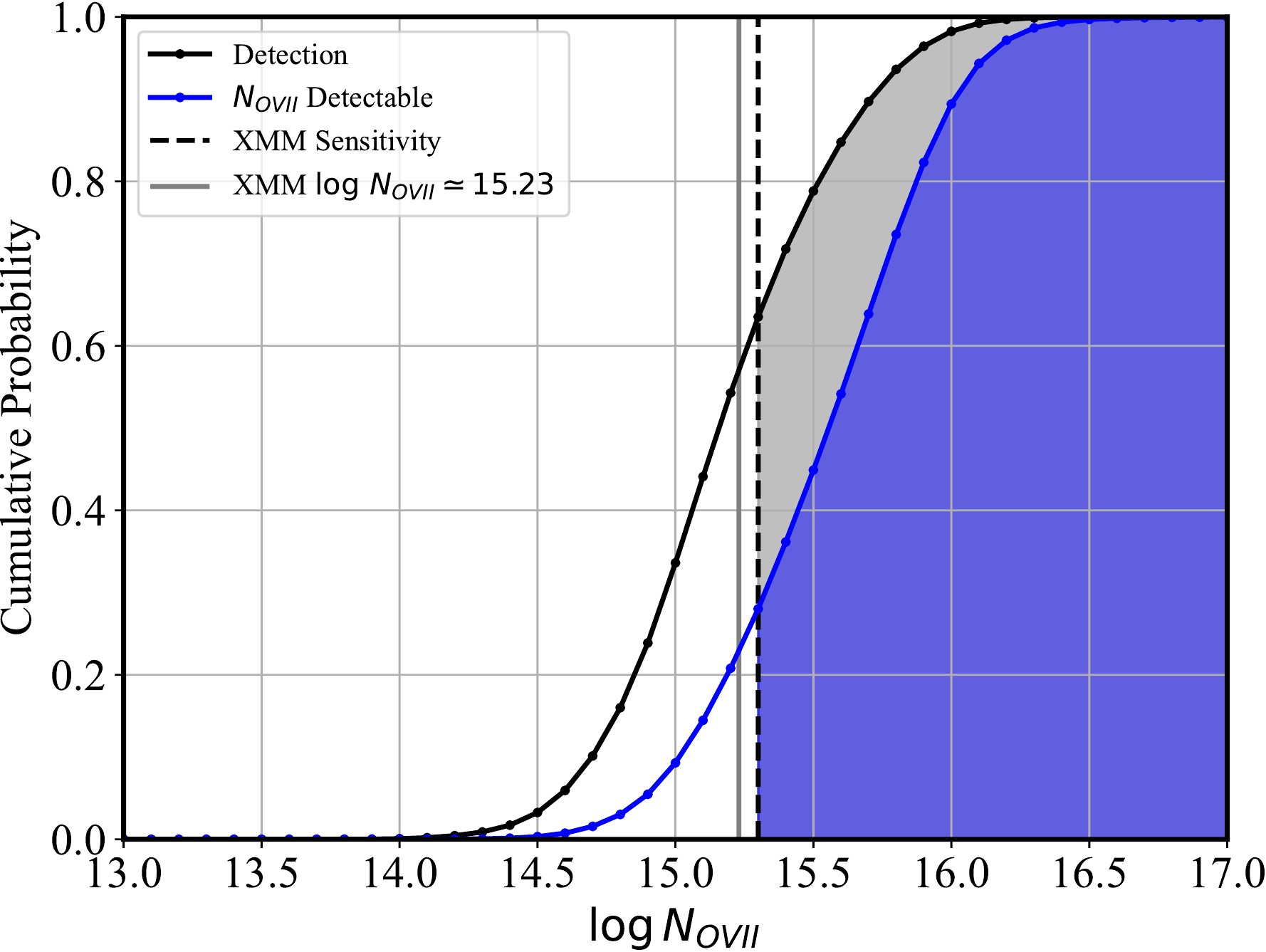}
    \includegraphics[width=3.2in]{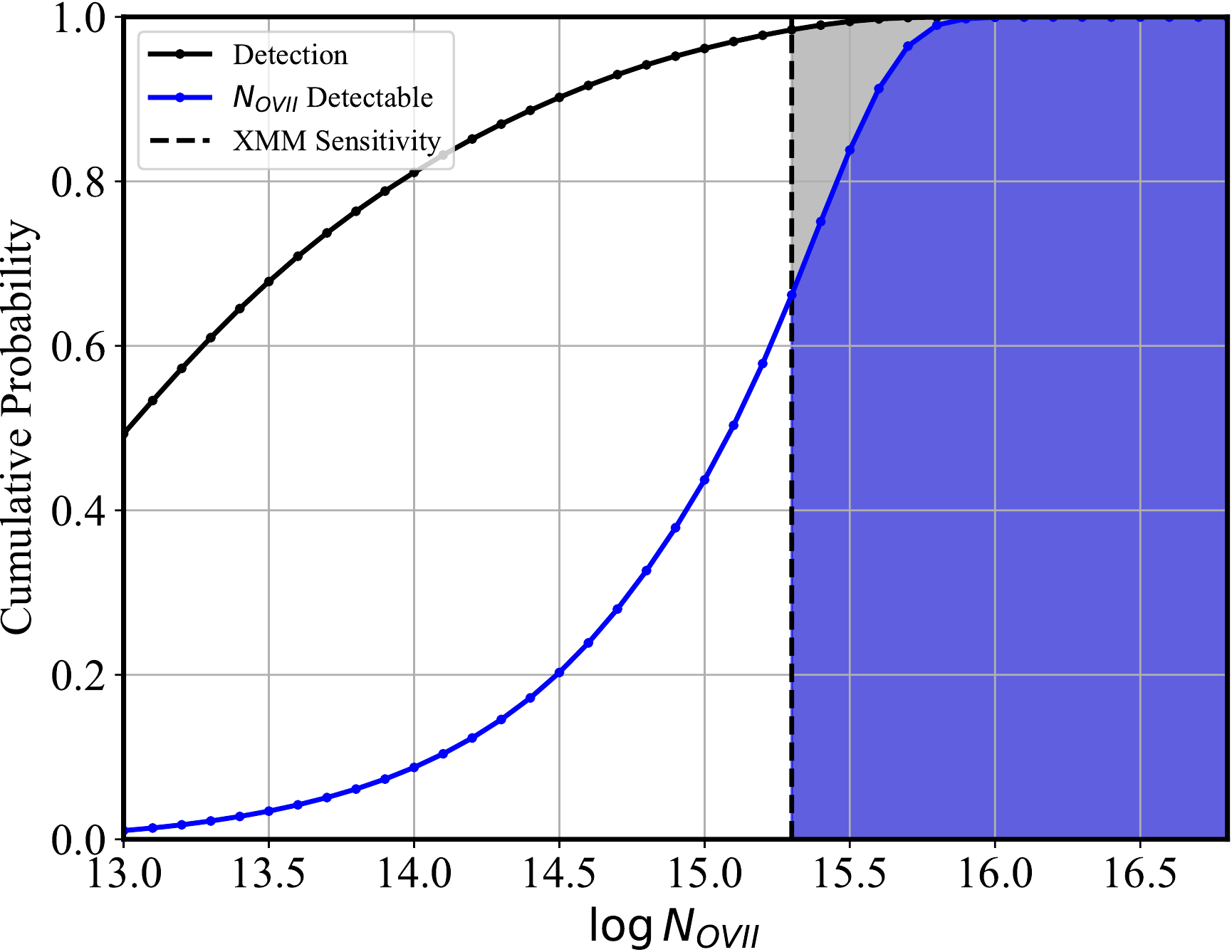} 
    \caption{(a) Cumulative distributions for \ovii\ with \ovi\ prior, and (b) for \ovii\ with \hi\ priors. 
    In both panels, black is the cumulative probability distributions for the possible 
    detection of an \ovii\ line, 
    for a fixed value of \ovi\ as in Fig.~\ref{fig:conditional}. In blue, for the same distributions, is the cumulative
    fraction $P_j$ of the column density of \ovii\ 
    above a given threshold, according to \eqref{eq:Pi}. On the left panel, for a sensitivity
    of approximately $\log N=15.3$ (15.1) for \ovii, \eagle\ 
    predicts that a sightline with $\log N_{\text{OVI}}=14.1$ is expected to have
    a probability $p=0.36$ (0.56) to intercept a sightline with $\log N_{\text{OVII}} \geq 15.3$ (15.1), and it is  is sensitive to 
    a fraction $P_j \simeq 0.72$ (0.86) of \ovii.} 
    \label{fig:bias}
\end{figure*}

\section{Discussion and Conclusions}
\label{sec:discussion}

This paper has presented a comprehensive statistical framework for the analysis of X--ray absorption lines
from the WHIM, with the goal of drawing inferences on the associated cosmological density of the WHIM, $\Omega_{\text{WHIM,X}}$.
The method is based on the search of X--ray absorption lines at fixed redshift where prior FUV observations
have identified possible absorption from the lower--temperature WHIM, and the use of
joint distribution functions of FUV and X--ray ions based on cosmological simulations.

Within this framework, we have searched the \es\ \xmm\ spectra for prominent
resonant absorption lines from the intervening WHIM. The method we have used is
based on the search near 8 fixed redshifts where FUV lines, namely \ovi\ and \hi\ BLAs,  had been previously detected 
by \cite{danforth2016}. 
The only FUV redshift system with evidence for a WHIM resonant absorption line
is $z=0.1876$, where a possible \ovii\ absorption line model is preferred by the data 
with a 99\% level of confidence, approximately corresponding to a 2.6--$\sigma$ significance. 
Give the limited statistical significance, this detection should be regarded as tentative. Accordingly,  we did not attempt
any modelling of the physical characteristics of the potential FUV/X--ray absorber, but instead
parameterized the associated cosmological inferences as a function
of temperature and abundance. If this signal is interpreted as a detection of \ovii\ from the WHIM,
such a detection would correspond to a significant fraction ($\OWHIMX/\Omega_b=0.44\pm0.22$) of the cosmological
density of the present--day baryons. 

There is no statistically significant evidence of any other 
X--ray ions along the sightline. To further examine the possibility of WHIM absorption, we performed a stacking of the data for the
two \ovii\ and \oviii\ absorption lines following the methods of Sect.~\ref{sec:stack}. The results of this analysis are reported
in Table~\ref{tab:stack} and in Figure~\ref{fig:oviistack}.
For \ovii\ from the entire sample of seven independent systems, the significance of detection is at 
a similar confidence level as  redshift system 1 \ovii\ alone. We therefore conclude that there is only marginal evidence for 
\ovii\ absorption along the sightline to \es.
For \oviii, there is no overall evidence for absorption lines from the
stacking of the spectra.

Given the limited statistical significance in the detection of just one \ovii\ system, 
we also provided an overall upper limit to $\OWHIMX$ based on
the systematic non--detection of WHIM
along the sightline to \es. This upper limit is based on a statistical study of the distribution of
X--ray and FUV ions in \texttt{EAGLE} \cite{wijers2019}, and yields a value of $\OWHIMX/\Omega_b \leq 0.90$,
assuming 100\% \ovii\ fraction, and a chemical abundance of 10\% Solar for oxygen. The limit becomes less
stringent if the WHIM systems are at a significantly lower temperature, where the ionization fraction of 
\ovii\ is significantly smaller.

Both the present analysis, and that of
\cite{kovacs2019}, highlight the importance and the potential of using FUV priors to identify faint X--ray absorption lines,
despite the overall difficulty of this task with the current generation of X--ray grating spectrometers.
In fact, the use of FUV priors enables the stacking of data at predetermined redshifts with the 
ensuing increase in sensitivity to absorbing columns, and improves the identification of faint absorption features
such as the possible \ovii\ absorption line at $z=0.1876$ identified in this paper.

It is useful to address the question of whether
the X--ray absorbing WHIM along the sightline towards \es\ is consistent with the expectation
from the \texttt{EAGLE} simulations. Our analysis, largely based on an earlier study by \cite{wijers2019},
uses the joint distribution of \ovi\ and \ovii\ column densities, and that of \hi\ and \ovii\
(see Fig.~\ref{fig:joint}). Given the priors provided by the FUV detections,
the probability of an \ovii\ detection at one of the three independent redshifts examined was estimated as 
$p\simeq 0.3$, depending on the exact value of the assumed sensitivity of the data (see Fig.~\ref{fig:bias}).
Since we have at most just one possible detection, and with limited statistical significance, our \xmm\ data are
clearly consistent with the expectation. Specifically, the probability of no detections based on a 
binomial distribution with $p=0.3$ and $N=3$ tries is 34\%, and the probability of one detection is 19\%.
For the four redshift systems with \hi\ BLA priors, the probability of an \ovii\ line above the sensitivity is smaller,
typically of order only a few percent. The systematic non--detection of \ovii\ at these redshifts is again
consistent with the \texttt{EAGLE} predictions. The main reasons for the  low probability of detection of
X--ray absorption are 1) the relatively low quality of the X--ray data limiting the detectable column density level and 2) the relatively short redshift path length probed rendering the number of absorbers low. The currently best X-ray data are only sensitive to \ovii\ column density well in 
excess of $10^{15}$~cm$^{-2}$, while the FUV data are able to detect column densities of order $10^{13}$~cm$^{-2}$.
In fact, according to \cite{wijers2019} (see discussion in their Sect.~4.1), \eagle\ predicts less than one
\ovii\ system for the present \es\ data.
This in turn indicates that future searches for the X--ray absorbing WHIM require substantially 1) more sensitive
instruments, such as the ones being developed for the future missions \emph{Athena} \citep[e.g.][]{barret2020}, 
\emph{Arcus} \citep{smith2019}, \emph{Lynx} \citep{gaskin2019} or \emph{HUBS} \citep{cui2020}, and 2) larger path lengths so that the number of intervening absorbers is feasible.



In addition to the search at the FUV redshifts, we also performed a dedicated analysis near two wavelengths where
\cite{nicastro2018} tentatively identified \ovii\ absorption from a blind search
of the entire \es\ RGS spectra. 
For the  first wavelength of $\lambda=30.98$\AA, the spectra
(see bottom panel of Fig.~\ref{fig:nicastro}) do confirm the presence of an absorption line feature with high
statistical significance, even after accounting for the redshift trials in the blind search. 
For the other feature near the wavelength $\lambda=29.28$\AA, 
we find no statistically significant evidence of absorption (see upper panel of
Fig.~\ref{fig:nicastro}), consistent with the re--analysis done
by the same group \citep[][]{nicastro2018b}, where the initial report of a detection
at that redshift was reanalyzed and its significance reduced. 
The confirmed \cite{nicastro2018} absorption line, if identified as \ovii\ He--$\alpha$
at $z=0.4339$, would imply a cosmological density of baryons that is indeed consistent with the
hypothesis that the missing baryons are in the hot phase of the intergalactic WHIM,
for an estimated $\OWHIMX/\Omega_b=0.65\pm0.11$, assuming typical physical conditions for
the absorber. However, \cite{johnson2019} showed that \es\ is likely a member of a galaxy
group at $z=0.433$, and \cite{dorigo2022} argues that the most likely redshift
for \es\ corresponds with that of the X--ray absorber. It is therefore likely that the serendipitous \cite{nicastro2018}
absorption line is intrinsic to the source or associated with the
group environment, rather than the WHIM.

\section*{Data Availability}
The \xmm\ and \chandra\ data analyzed in this paper are available from
the public NASA W3Browse archive (https://heasarc.gsfc.nasa.gov/cgi-bin/W3Browse/w3browse.pl).
Processed data such as spectra and response functions can be obtained by the authors upon request.

\section*{Acknowledgments}
Massimiliano Bonamente and David Spence acknowledge support from NASA through an ADAP grant awarded to the University of Alabama in Huntsville for the project 
`Closing the Gap on the Missing Baryons at Low Redshift with
multi–wavelength observations of the Warm–Hot Intergalactic Medium'.
Nastasha Wijers was supported by a CIERA Postdoctoral Fellowship.

\bibliographystyle{mnras}
\input{main.bbl}

\end{document}